\documentclass[journal,twoside,web]{ieeecolor}
\pdfoutput=1
\usepackage{generic}
\usepackage{cite}
\usepackage{amsmath,amssymb,amsfonts}
\usepackage{algorithmic}
\usepackage{graphicx}
\usepackage{algorithm,algorithmic}
\usepackage{mathtools,mathbbol,dsfont}
\usepackage{bm}
\usepackage{textcomp}
\usepackage[all]{xy}
\usepackage{chemarrow}
\usepackage{color}
\usepackage{mathrsfs,array}
\makeatletter

\newcommand{\Rmnum}[1]{\expandafter\@slowromancap\romannumeral #1@}
\makeatother
\usepackage{textcomp}
\usepackage{booktabs}
\usepackage[version=4]{mhchem}

\newtheorem{theorem}{Theorem}[section]
\newtheorem{corollary}{Corollary}[section]
\newtheorem{lemma}{Lemma}[section]
\newtheorem{definition}{Definition}[section]
\newtheorem{proposition}[theorem]{Proposition}
\newtheorem{remark}{Remark}[section]
\newtheorem{example}{Example}[section]

\def\BibTeX{{\rm B\kern-.05em{\sc i\kern-.025em b}\kern-.08em
    T\kern-.1667em\lower.7ex\hbox{E}\kern-.125emX}}
\begin{document}
\title{Input-to-state stability-based chemical reaction networks composition for molecular computations}
\author{Renlei Jiang, Yuzhen Fan, Di Fan, Chuanhou Gao, \IEEEmembership{Senior Member, IEEE}, and Denis Dochain 
\thanks{This work was funded by the National Nature Science Foundation
of China under Grant No. 12320101001.} 
\thanks{R. Jiang, Y. Fan, D. Fan and C. Gao are with the School of Mathematical Sciences, Zhejiang University, Hangzhou, China (e-mail: jiangrl@zju.edu.cn, yuzhen$\_$f@zju.edu.cn, fandi@zju.edu.cn, gaochou@zju.edu.cn (correspondence)). }
\thanks{D. Dochain is with 
ICTEAM, UCLouvain, B ˆatiment Euler, avenue Georges Lemaˆıtre 4-6, 1348 Louvain-la-Neuve, Belgium (e-mail: denis.dochain@uclouvain.be).}}

\maketitle

\begin{abstract}
Molecular computation based on chemical reaction networks (CRNs) has emerged as a promising paradigm for designing programmable biochemical systems. However, the implementation of complex computations still requires excessively large and intricate network structures, largely due to the limited understanding of composability, that is, how multiple subsystems can be coupled while preserving computational functionality. Existing composability frameworks primarily focus on rate-independent CRNs, whose computational capabilities are severely restricted. This article aims to establish a systematic framework for composable CRNs governed by mass-action kinetics, a common type of rate-dependent CRNs. Drawing upon the concepts of composable rate-independent CRNs, we introduce the notions of mass-action chemical reaction computers (msCRCs), dynamic computation and dynamic composability to establish a rigorous mathematical framework for composing two or more msCRCs to achieve layer-by-layer computation of composite functions. Further, we derive several sufficient conditions based on the notions of input-to-state stability (ISS) to characterize msCRCs that can be composed to implement desired molecular computations, thereby providing theoretical support for this framework. Some examples are presented to illustrate the efficiency of our method. Finally, comparative results demonstrate that the proposed method exhibits notable advantages in both computational ability and accuracy over the state-of-the-art methods.

\end{abstract}


\section{Introduction}
\IEEEPARstart{M}{olecular} computation aims to perform computations at the molecular scale by harnessing reactions between biological molecules. Recently, there has been a growing interest in this field partly due to the fast development of synthetic biology, especially in the design of biochemical elements and their assembly to perform sophisticated calculations \cite{Chen_2024_synthetic}. Compared to silicon-based hardware computation, limited by physical factors, biocomputing is considered a better substitute for the fields of medical diagnosis\cite{Danino_2015_programmable}, energy\cite{Peralta_2012_microbial}, and so on.

Chemical reaction network (CRN) is the main carrier for molecular computation. Typically, it characterizes the interaction of chemical species through reactions, and is commonly used to represent complex biochemical processes, such as metabolism, gene regulation, etc. Mathematically, a CRN is usually modeled by a group of ordinary differential equations (ODEs) under mass-action kinetics, often called mass-action system (MAS) that capture the changes of concentrations of each species over time. The structure and parameters (reaction rate constants) determine the dynamic behavior of the MAS, often exhibiting stability \cite{Feinberg_1987_chemical}, persistence \cite{Craciun_2013_persistence}, oscillations \cite{Domijan_2009_bistability}, and bifurcations \cite{Conradi_2007_saddle}. This versatility renders them a potent tool for understanding both natural and engineered chemical systems, as well as a powerful framework for molecular computation.

The pioneering work in this field is to utilize the inherent parallelism of chemical reactions for solving the Hamiltonian path problem \cite{Adleman_1994_molecular}, which is a well-known NP-hard problem. Subsequently, some research focused on using CRNs to implement Boolean operations \cite{Arkin_1994_computational} and elementary algebraic operations \cite{Buisman_2009_computing}, such as addition, subtraction, multiplication, and division. Further, Fages \textit{et al.} proved that MASs are Turing universal \cite{Fages_2017_strong}, 
which means that any functional computation can be embedded into a class of polynomial ODEs and implemented by CRNs. Meanwhile, it was shown that any MAS could be physically implemented through DNA strand displacement \cite{Soloveichik_2010_dna}, indicating that abstract CRN models can be always instantiated in terms of tangible molecular entities. These studies were attracting more researchers towards developing methods to embed learning and reasoning capabilities into chemical reaction systems. On the one hand, there were some tangible implementations of molecular computing systems through DNA strand displacement reactions in experimental wet labs \cite{Cherry_2018_scaling,Okumura_2022_nonlinear,Chen_2024_synthetic}. On the other hand, theoretical research was also undergo rapid development. Vasic \textit{et al.} proposed a novel language, termed CRN++, for programming deterministic CRNs to perform computation \cite{Vasic_2020_crn++}, which contributes to establishing a comprehensive framework for molecular programming. Chou \textit{et al.} \cite{Chou_2017_chemical} proposed a method for achieving logarithmic computation using CRNs with a low number of reactions, whose precision can be enhanced by adjusting the parameters. Chen et. al. \cite{Chen_2014_rate} used a special type of CRN, known as rate-independent CRNs, to implement computation, and proved that a function is computable by such CRNs if and only if it is continuous piecewise linear. \cite{Chalk_2019_composable} further explored the composability of these CRNs, revealing that the functions computed by composable rate-independent CRNs possess superadditive properties. Samaniego \textit{et al.} \cite{Samaniego_2020_sequestration} proposed a motif for performing approximate derivative operations by molecular sequestration and delays, while also discussing its potential biological implementations. Anderson \textit{et al.} \cite{Anderson_2025_chemical} viewed CRNs as analog computers and investigated the speed of computation carried out by CRNs. In addition, several studies explored the implementation of neural networks using CRNs \cite{Vasic_2022_programming,Anderson_2021_reaction,Fan_2025_automatic}.

While CRN-based molecular computations have achieved remarkable results, the field still faces a fundamental challenge arising from the inherent conflict between the parallel nature of chemical reactions and the sequential execution of computations. A classical solution to this issue is the utilization of chemical oscillators, which is a set of chemical reactions admitting a certain oscillatory behavior \cite{Lachmann_1995_computationally,Shi_2025_controlling}. This oscillatory behavior can generate symmetrical clock signals, which segments the time interval $T$ into a series of phases with equal length. In each phase, only one species has a concentration strictly larger than $0$ while the concentrations of other species are nearly $0$. Consequently, it possesses the capability to catalyze specific reactions while deactivating others, which facilitates the segregation of two or more concurrent reactions and orchestrates their sequential occurrence. However, for a complex multi-step molecular computation task (for example, the automatic implementation of neural networks in \cite{Fan_2025_automatic}), each step requires a phase to accomplish the computation. This necessity leads to the requirement of an extensive number of phases for the entire computational task (32 phases in \cite{Fan_2025_automatic}), which is rarely observed in both natural and synthetic biological systems. In addition, the utilization of chemical oscillators introduces computational errors. The first type of error arises from the finite length of each phase, which prevents reactions from reaching equilibrium \cite{Fan_2024_automatic}. The second type of error occurs due to the incomplete elimination of minor signaling species in the symmetrical clock signals, resulting in the inability to fully deactivate certain reactions \cite{Shi_2025_controlling}. These errors accumulate as the number of phases increases, ultimately leading to significant deviations in the computational results. Although there was a study on the composability of rate-independent CRNs \cite{Chalk_2019_composable}, it is difficult to apply in general computations due to the inherent limitations in the computational power of rate-independent CRNs \cite{Chen_2014_rate}. 
Although studies on the composability of rate-independent CRNs have been conducted \cite{Chalk_2019_composable}, their application to general computations is limited due to inherent constraints on the computational power of rate-independent CRNs \cite{Chen_2014_rate}.

To address the above issues, it is imperative to investigate the composability of multiple simple molecular computational systems, specifically, to explore the possibility of leveraging their composition to implement the corresponding composition computation. In this paper, we adapt the definition of composability from rate-independent CRNs \cite{Chalk_2019_composable} to MASs, and study the condition for MASs to be composable in implementing computations. The main contributions are listed below: 

\begin{itemize}
    \item We extend the concepts of composability to MASs, and investigate its crucial role in computing composite functions (\textit{Theorem} \ref{main_thm} and \textit{Corollary} \ref{coro_thm1}).
    \item We propose several sufficient conditions that allow two or more MASs to be composable (\textit{Proposition} \ref{prop_simple_structure}, \textit{Theorem} \ref{Thm_ISS_net}, \textit{Corollary} \ref{coro_thmISS} and \textit{Proposition} \ref{iss_condition}) leveraging input-to-state stability (ISS) \cite{Sontag_1989_smooth}.
    \item We compare our method with state-of-the-art approaches, revealing its potential for computing complex functions accurately without adding catalyst species.
\end{itemize}

This paper is organized as follows. Some preliminaries on CRN and dynamic computation are given in Section \ref{Section_2}. In Section \ref{Section_3}, we present the motivation for our study, followed by a detailed analysis of mass-action chemical reaction computers (msCRCs) composition and dynamic composability based on ISS, which can be leveraged to achieve layer-by-layer molecular computation for composite functions. Several illustrative examples are also provided in this section. In Section \ref{Section_4}, we compare our method with state-of-the-art approaches and highlight the former's advantages compared with the latter. Finally, Section \ref{Section_5} is dedicated to the conclusion of the whole paper.

\textbf{Notations}: $\mathbb{R}^n,\mathbb{R}_{\ge0}^n,\mathbb{R}_{>0}^n,\mathbb{Z}_{\ge 0}^n$ denote $n$-dimensional real space, nonnegative space, positive space and nonnegative integer space, respectively. $x^{v_{\cdot j}}$ represents $\prod_{i=1}^nx_i^{v_{ij}}$, where $x \in \mathbb{R}^n$, $v_{\cdot j}\in \mathbb{Z}^n$ and $0^0=1$. The norm $\left \|\cdot \right \|$ is regarded as $\infty$-norm in Euclidean space. A class $\mathcal{K}$ function $\gamma:\mathbb{R}_{\ge 0} \to \mathbb{R}_{\ge 0}$ is continuous strictly increasing and satisfies $\gamma(0)=0$; it is a class $\mathcal{K}_{\infty}$ function if in addition $\gamma(s) \to \infty$ as $s \to \infty$. A class $\mathcal{KL}$ function $\beta(s,t):\mathbb{R}_{\ge 0} \times \mathbb{R}_{\ge 0} \to \mathbb{R}_{\ge 0}$ satisfies that for each fixed $t$ the mapping $\beta(\cdot,t)$ is class $\mathcal{K}$ function and for each fixed $s$ it decreases to zero on $t$ as $t \to \infty$.

\section{CRN concepts and dynamic computation}\label{Section_2}
In this section, we will present some basic concepts related to CRN \cite{Feinberg_1987_chemical} and the computation of MAS \cite{Chalk_2019_composable}. 

\subsection{Chemical Reaction Network}
Consider a CRN with $n$ species, denoted by $S_1,...,S_n$, and $r$ reactions with the $j$th reaction $\mathcal{R}_j$ written as
$$\sum_{i=1}^nv_{ij}S_i\to \sum_{i=1}^nv'_{ij}S_i,$$
where $v_{.j}, v'_{.j}\in\mathbb{Z}_{\geq 0}^n$ represent the complexes of reactant and resultant of $\mathcal{R}_j$, respectively. For simplicity, this reaction is often written as $v_{.j}\to v'_{.j}$. We thus present the related notions in the CRN theory.  

\begin{definition}[CRN]
    A CRN consists of three finite sets: 
    \begin{enumerate}
        \item a finite \textit{species} set $\mathcal{S}=\{ S_1,...,S_n\}$;
        \item a finite \textit{complex} set $\mathcal{C}=\bigcup_{j=1}^r{\left\{ v_{\cdot j},v_{\cdot j}^{\prime} \right\}}$, where the $i$th entry of $v_{\cdot j}$, i.e., $v_{ij}$, represents the stoichiometric coefficient of species $S_i$ in complex $v_{\cdot j}$;
        \item a finite \textit{reaction} set $\mathcal{R}=\bigcup_{j=1}^r{\left\{ v_{\cdot j}\rightarrow v_{\cdot j}^{\prime} \right\}}$ satisfying
        \begin{enumerate}
            \item $\forall v_{\cdot j}\rightarrow v_{\cdot j}^{\prime}\in \mathcal{R}, v_{\cdot j}\ne v_{\cdot j}^{\prime}$,
            \item $\forall v_{\cdot j} \in \mathcal{C}, \exists v_{\cdot j}^{\prime} \in \mathcal{C}$ such that $v_{\cdot j}\rightarrow v_{\cdot j}^{\prime}\in \mathcal{R}$ or $v_{\cdot j}^{\prime}\rightarrow v_{\cdot j}\in \mathcal{R}$.
        \end{enumerate}
    \end{enumerate}
 The triple $(\mathcal{S},\mathcal{C},\mathcal{R})$ is usually used to express a CRN.
\end{definition}

\begin{definition}[Stoichiometric subspace]
    For a $(\mathcal{S},\mathcal{C},\mathcal{R})$, the linear subspace $\mathscr{S}=\textrm{span}\{ v_{\cdot 1}^{\prime}-v_{\cdot 1},...,v_{\cdot r}^{\prime}-v_{\cdot r}\}$ is called the \textit{stoichiometric subspace} of the network.
\end{definition}

\begin{definition}[Stoichiometric compatibility class]\label{def_stoichiometric compatibility class}
For a $(\mathcal{S},\mathcal{C},\mathcal{R})$, let $s_0 \in \mathbb{R}_{\ge0}^n$, the set $s_0+\mathscr{S}=\{s_0+s|s \in \mathscr{S}\}$ is called the \textit{stoichiometric compatibility class} of $s_0$. Further, $(s_0+\mathscr{S}) \bigcap \mathbb{R}_{\ge 0}^n$ is called the \textit{nonnegative stoichiometric compatibility class} of $s_0$, and $(s_0+\mathscr{S}) \bigcap \mathbb{R}_{> 0}^n$ is called the \textit{positive stoichiometric compatibility class} of $s_0$.
\end{definition}

\begin{example}\label{example1}
A CRN follows
\begin{align}\label{additionCRN}
       S_{1} &\longrightarrow S_{1} + S_{2}\ , &   
        S_{3} &\longrightarrow S_{3} + S_{2}\ ,  &           S_{2} &\longrightarrow \varnothing,  
    \end{align}
 where the last reaction refers to an outflow reaction. The species set is $\{S_1,S_2,S_3\}$, the complex set is $\{(1,0,0)^\top,(1,1,0)^\top,(0,0,1)^\top,(0,1,1)^\top,(0,1,0)^\top\}$, and the stoichiometric subspace is span$\{(0,1,0)^\top,(0,-1,0)^\top\}$. Note that species $S_1,~S_3$ do not change during the reaction process. We call species exhibiting such a property in a reaction a \textit{catalyst}. 
\end{example}

When a CRN is equipped with mass-action kinetics, the reaction rate is measured according to the power law with respect to the species concentrations. For example, the reaction rate of $v_{\cdot j} \rightarrow v_{\cdot j}^{\prime}$ is evaluated by
\begin{equation}\label{kinetic}
   \kappa_j(s)=k_js^{v_{\cdot j}}, 
\end{equation}
where $k_j >0$ represents the rate constant, and $s \in \mathbb{R}_{\ge 0}^n$ with each element $s_i~(i=1,...,n)$ to represent the concentration of the species $S_i$. The reaction $\mathcal{R}_j$ is accordingly labeled by $v_{\cdot j} \overset{k_j}{\rightarrow} v_{\cdot j}^{\prime}$. 
\begin{definition}[MAS]\label{def_MAS}
     A \textit{MAS} is a CRN $(\mathcal{S},\mathcal{C},\mathcal{R})$ taken with a mass-action kinetics $\kappa$, often labeled by $(\mathcal{S},\mathcal{C},\mathcal{R},\kappa)$.
\end{definition}

The dynamics of $(\mathcal{S},\mathcal{C},\mathcal{R},\kappa)$ describing the change of concentrations of all species over time $t$ thus follows
\begin{equation}\label{general dynamics}
    \frac{\mathrm{d}s(t)}{\mathrm{d}t}=\Gamma \kappa (s),
\end{equation}
where $\Gamma \in \mathbb{Z}^{n \times r}$ is the stoichiometric matrix with the $j$th column $\Gamma_{\cdot j}=v_{\cdot j}^{\prime}-v_{\cdot j}$ to represent the reaction vector of $\mathcal{R}_j$, and $\kappa (s)$ is the vector-valued function from $\mathbb{R}^n_{\geq 0}$ to $\mathbb{R}^r_{\geq 0}$ with the $j$th element $\kappa_j (s)$ constrained by (\ref{kinetic}). The dynamics of (\ref{general dynamics}) is essentially polynomial ODEs. By integrating (\ref{general dynamics}) from $0$ to $t$, we get
\begin{equation}\label{integ_dynamics}
 s(t) = s_0 + \sum_{j=1}^r \Gamma_{\cdot j} \int_0^t \kappa_j(s(\tau))\mathrm{d}\tau,   
\end{equation}
where $s_0=s(0)$ is the initial state of $(\mathcal{S},\mathcal{C},\mathcal{R},\kappa)$. It is clear that the state of this MAS will evolve in the nonnegative stoichiometric compatibility class of $s_0$, i.e., in $(s_0+\mathscr{S}) \bigcap \mathbb{R}_{\ge 0}^n$. 

We also use the CRN in $(\ref{additionCRN})$ to exhibit the
stoichiometric matrix to be $\Gamma=\begin{pmatrix}
  0&0  &0 \\
  1& 1 & -1\\
  0& 0 &0
\end{pmatrix}$, the reaction rate function to be $r(s)=(s_1,s_3,s_2)^{\top}$, and the dynamics to be 
 \begin{align}\label{ex1_dynamics}
\frac{\mathrm{d} s_{2}}{\mathrm{d} t} &= s_{1} + s_{3} - s_{2}\ , & 
\frac{\mathrm{d} s_{1}}{\mathrm{d} t} &=\frac{\mathrm{d} s_{3}}{\mathrm{d} t} = 0 
\end{align}

\begin{definition}[Equilibrium]
    For a $(\mathcal{S},\mathcal{C},\mathcal{R},\kappa)$ governed by (\ref{general dynamics}),
    \begin{enumerate}
        \item if a constant vector $\bar{s} \in \mathbb{R}_{\ge 0}^n$ satisfies $\Gamma \kappa(\bar{s})=0$, then $\bar{s}$ is called a \textit{non-negative equilibrium point};
        \item if a set $\mathcal{A} \subset \mathbb{R}_{\ge 0}^{n}$ satisfies $\forall s \in \mathcal{A}, \Gamma \kappa(s)=0$, then $\mathcal{A}$ is called a \textit{equilibrium set}.
    \end{enumerate}
    Both are collectively known as \textit{equilibrium}.
\end{definition}

\begin{definition}[Stability \cite{Rouche_1977_stability}]\label{stability}
    The equilibrium point $\bar{s}$ of system (\ref{general dynamics}) is
    \begin{enumerate}
        \item stable if for any $\varepsilon >0$, there is $\delta=\delta(\varepsilon)$ such that
        $$\left \|s(0)-\bar{s} \right \|<\delta  \Rightarrow \left \|s(t)-\bar{s} \right\|<\varepsilon, \quad \forall t \ge 0;$$
        \item asymptotically stable if it is stable and $\delta$ can be chosen such that
        $$\left \|s(0)-\bar{s} \right \|<\delta  \Rightarrow \lim_{t \to \infty}s(t)=\bar{s};$$
        \item globally asymptotically stable if 
        $$\lim_{t \to \infty}s(t)=\bar{s}, \quad \forall s(0) \in \mathbb{R}_{\ge 0}^n;$$
        \item exponentially stable if there are positive constants $M,\lambda,r>0$ such that
        $$\left \|s(0)-\bar{s} \right \| \le r \Rightarrow \left \|s(t)-\bar{s} \right \| \le Me^{-\lambda t}, \quad \forall t \ge 0.$$
    \end{enumerate}
\end{definition}

We also give the definition of ISS that will be used later. 
\begin{definition}[ISS \cite{Sontag_1989_smooth}]
Assume given a control system  
\begin{equation}\label{general_system}
    \dot{s}=f(u,s), \quad s(0)=s_0,
\end{equation}
where $f:\mathbb{R}^m \times \mathbb{R}^{n-m} \to \mathbb{R}^{n-m}$ is locally Lipschitz in $u$ and $s$, and the input $u(t)$ is a piecewise continuous, bounded function of $t$ for all $t \ge 0$. Also, for a fixed $\bar{u}\in\mathbb{R}^m$ suppose $\bar{s}$ is a globally asymptotically stable equilibrium point in (\ref{general_system}) evaluated at $u=\bar{u}$. Then the system (\ref{general_system}) is said to be ISS regarding $(\bar{u},\bar{s})$ if there exist a class $\mathcal{KL}$ function $\beta$ and a class $\mathcal{K}$ function $\gamma$ such that for any initial state $s_0$ and any bounded input $u(t)$, the solution $s(t)$ exists for all $t \ge 0$ and satisfies
\begin{equation}\label{ISS}
    \left \| s \left( t \right) -\bar{s} \right \| \le \beta \left( \left\| s_0 - \bar{s} \right\|, t \right)+ \gamma( \sup \limits_{0\le \tau \le t}\left\| u \left( \tau \right) -\bar{u} \right\| ).
\end{equation}
\end{definition}

Note that the function $f$ here and also those $f,g,...$ appearing later are all assumed to be locally Lipschitz in their arguments. We will not assert this specially in the subsequent locations. 

\subsection{The dynamic computation of MAS}
In this subsection, we leverage the work of Chalk \textit{et al.} \cite{Chalk_2019_composable}, and redefine the notions related to molecular computation.
\begin{definition}[msCRC]
    A msCRC is a tuple $\mathscr{C}=(\mathcal{S},\mathcal{C},\mathcal{R},\kappa,\mathcal{X},\mathcal{Y})$, where $(\mathcal{S,C,R},\kappa)$ is a MAS, $\mathcal{X} \subset \mathcal{S}$ is the input species set, and $\mathcal{Y}=\mathcal{S} \setminus \mathcal{X}$ is the output species set.
\end{definition}

\begin{definition}[Dynamic computation]\label{Def_dynamic_compute} Given a msCRC $\mathscr{C}=(\mathcal{S},\mathcal{C},\mathcal{R},\kappa,\mathcal{X},\mathcal{Y})$ with $m~(m<n)$ species in $\mathcal{X}$, and a positive function $\sigma:\mathbb{R}^{m}_{\ge 0} \to \mathbb{R}^{n-m}_{\ge 0}$, denote the dynamics of this msCRC by
    \begin{equation}\label{CRC_dynamics}
    \begin{cases}
        \dot{x}=f(x,y), \\
        \dot{y}=g(x,y),
    \end{cases} x(0)=x_0,y(0)=y_0,
    \end{equation}
where $x \in \mathbb{R}^{m}_{\ge 0}$ and $y \in \mathbb{R}^{n-m}_{\ge 0}$ are the concentrations of input species and output species, respectively. The msCRC is a dynamic computation of function $\sigma$, if the output species concentrations satisfy
    \begin{equation}\label{eq:dyCom}
  \lim_{t \to \infty}y(t)=\sigma(x_0).      
    \end{equation}
\end{definition}
We refer to $\lim_{t \to \infty}y(t)$ as the limiting steady state of the output $y(t)$ in the context.

Clearly, the dynamics of (\ref{CRC_dynamics}) is an alternative of the original form of (\ref{general dynamics}), which suggests that $f:\mathbb{R}_{\ge 0}^n \to \mathbb{R}_{\ge 0}^m,~g:\mathbb{R}_{\ge 0}^n \to \mathbb{R}_{\ge 0}^{n-m}$ have the structure of polynomical functions. In addition, the limit equality (\ref{eq:dyCom}) implies that the output species in the msCRC $\mathscr{C}=(\mathcal{S},\mathcal{C},\mathcal{R},\kappa,\mathcal{X},\mathcal{Y})$ should have a certain kind of stable behavior. Of course, the value of $\sigma(x_0)$ is not fixed, but depends on the structure of the function $\sigma(\cdot)$ and the parameter involved. Revisit \textit{Example} \ref{example1} and the corresponding dynamical equation (\ref{ex1_dynamics}), we then have $s_1(t)=s_1(0), s_3(t)=s_3(0)$, and 
$s_2(t)=s_1(0)+s_3(0)+\left (s_2(0)-s_1(0)-s_3(0) \right ) e^{-t}$. By setting input species set $\mathcal{X}=\{S_1,S_3\}$ and output species set $\mathcal{Y}=\{S_2\}$, we get $\lim_{t \to \infty}s_2(t)=s_1(0)+s_3(0)$. This means the msCRC induced by (\ref{additionCRN}) with $\mathcal{X}=\{S_1,S_3\}$ and $\mathcal{Y}=\{S_2\}$ is a dynamic computation of the function $\sigma(
  (s_1, s_3)^\top
)=s_1+s_3$. Essentially, the network given in (\ref{additionCRN}) can perform the addition computation between any two nonnegative real numbers in the form of $s_1(0)$ and $s_3(0)$. From the viewpoint of dynamic computation, the limiting steady state of the output is independent of its initial point $y_0$.

We give another example to dynamically compute a function that depends on the specific initial output concentration.
\begin{example}\label{ex_exp}
    A msCRC follows
    \begin{align}\label{exp_CRC}
      \quad X + Y &\overset{1}{\longrightarrow} X, &  
       X &\overset{1}{\longrightarrow} \varnothing  
    \end{align}
    with input species set $\mathcal{X}=\{X\}$ and output species set $\mathcal{Y}=\{Y\}$. It is easy to get
\begin{equation*}\label{function_ex1}
\begin{cases}
        \dot{x}=-x, \\
        \dot{y}=-xy;
    \end{cases} \Rightarrow \quad \begin{cases}
        x(t)=x_0e^{-t}, \\
        y(t)=y_0e^{-x_0\left(1-e^{-t}\right)}.
    \end{cases}
 \end{equation*}   
We thus have $\lim_{t \to \infty}y(t)=y_0e^{-x_0}$, which indicates that the msCRC is a dynamic computation of $\sigma(x)=y_0e^{-x}$. The initial output $y_0$  can be thought as the parameter in the model $\sigma(x)=y_0e^{-x}$. 
\end{example}

\section{Motivation statement and msCRCs Composition}\label{Section_3}
In this section, we formulate the motivation of the current study by computing the sigmoid function using CRNs, and further clarify the main method for controlling molecular computations accompanied by some theoretic supports. 

\subsection{Motivation statement}
\textit{Examples} \ref{example1} and \ref{ex_exp} realize the dynamic computation of addition and exponential function, respectively. Basically, they belong to the class of simple functions. However, when faced with composite functions, such as the sigmoid function $\sigma(x)=\frac{1}{1+e^{-x}}$, a common function used in algorithms related to neural network, one usually needs to cleverly design a msCSC so that it can dynamically compute the sigmoid function straightforwardly. It will pose a great challenge to do a msCRC design if the composite function considered is sophisticated. Another solution is by peeling the composite function to build several simple functions, and then to design the corresponding msCRCs layer by layer. Also for the sigmoid function, one can compute it through computing two functions $y=e^{-x}$ and $z=\frac{1}{1+y}$ in turn. Obviously, the layer-by-layer design of msCRCs is much easier than the direct design of msCRC according to a certain composite function. Moreover, it is a more general method, and has a potential to deal with any composite function.    

\begin{example}\label{ex_sigmoid}
 In this example, we compare two kinds of different computation patterns for the sigmoid function using CRNs. 
 
\textbf{Case 1): direct computation.} The following CRN \cite{Fan_2025_automatic} 
     \begin{equation}\label{dircom_sigmoid}
        X+Z \ce{<=>[1][1]}X+2Z, \quad X \overset{1}{\longrightarrow} \varnothing
    \end{equation}
has the dynamics and solution to be
\begin{equation*}
    \begin{cases}
        \dot{x}=-x, \\
        \dot{z}=xz-xz^2;
    \end{cases} \Rightarrow
    \begin{cases}
        x(t)=x_0e^{-t}, \\
        z(t)=\frac{1}{1+(\frac{1}{z_0}-1)e^{x_0(e^{-t}-1)}}.
    \end{cases}
\end{equation*}
By setting $\mathcal{X}=\{X\}$, $\mathcal{Y}=\{Z\}$ and $z_0=\frac{1}{2}$, we have
$$\lim_{t \to \infty}z(t)=\lim_{t \to \infty}\frac{1}{1+e^{-x_0(1-e^{-t})}}=\frac{1}{1+e^{-x_0}},$$ which means that the msCRC of (\ref{dircom_sigmoid}) can dynamically compute the sigmoid function directly. 
   
\textbf{Case 2): layer-by-layer computation.} In \textit{Example} \ref{ex_exp}, the given msCSC (\ref{exp_CRC}) can dynamically compute $y=e^{-x}$ if $y_0=1$ is set. Further, we construct 
 \begin{equation}\label{laycom_sigmoid}
Y + Z \overset{1}{\longrightarrow} Y,\quad Z \ce{<=>[1][1]} \varnothing,
\end{equation} 
whose dynamics and solution are
\begin{equation*}\label{function_ex2}
    \begin{cases}
        \dot{y}=0, \\
        \dot{z}=1-z-yz;
    \end{cases} \Rightarrow
    \begin{cases}
        y(t)=y_0, \\
       z(t)=\frac{1}{1+y_0}+\left(z_0-\frac{1}{1+y_0}\right)e^{-\left (1+y_0 \right)t}.
    \end{cases}
\end{equation*} 
Set $\mathcal{X}=\{Y\}$, $\mathcal{Y}=\{Z\}$ and $z_0\geq 0$, then we get the msCRC in (\ref{laycom_sigmoid}) can dynamically compute the function $z=\frac{1}{1+y}$.
\end{example}

\begin{figure}
    \centering
    \includegraphics[width=0.47\textwidth]{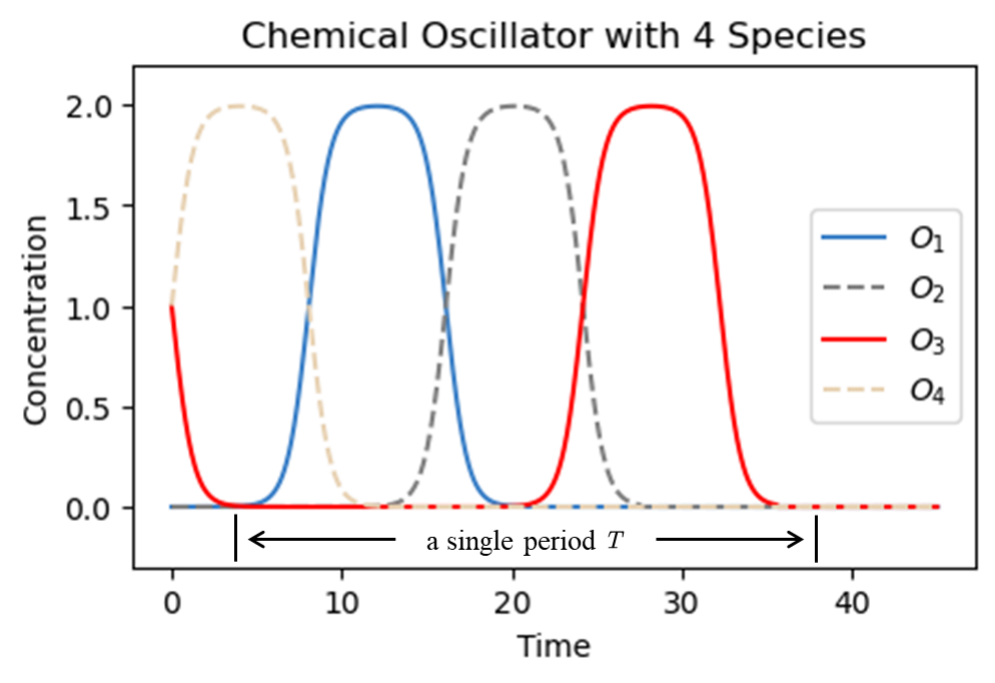}
    \caption{A schematic diagram on chemical oscillator with four species, in which $O_1$ and $O_3$ are used to control two reaction networks.}
    \label{fig_oscillator}
\end{figure}

Note that the reactions in (\ref{exp_CRC}) and (\ref{laycom_sigmoid}) need to take place in turn, then they can produce layer-by-layer dynamics to finish the corresponding computation. However, if these two networks are directly put together, the dynamics will change due to the coupling between them. The new dynamics is
 \begin{equation}\label{change_dynamics}
        \begin{cases}
        \dot{x}=-x, \\
        \dot{y}=-xy, \\
        \dot{z}=1-(1+y)z, \\
        \end{cases}
        \end{equation}    
which clearly violates the expected layer-by-layer dynamics (Of course, the coupled dynamics can still work for computing the sigmoid function through our current analysis method. We will prove this point in the subsequent section. At present, we expect to obtain the layer-by-layer dynamics according to the composition rule of the sigmoid function). To control the reactions taking place in turn, Lachmann and Sella \cite{Lachmann_1995_computationally} proposed to use oscillating signals as catalysts, which are generated by 
\begin{equation}\label{oscillator}
\begin{split}
    O_i+O_{i+1} &\overset{k_0}{\longrightarrow} 2O_{i+1}~(1\le i \le r-1), \\
    O_r+O_1 &\overset{k_0}{\longrightarrow} 2O_1,
\end{split}
\end{equation}
with dynamics
\begin{equation}
    \begin{cases}
        \dot{o}_1=k_0(o_1o_r-o_1o_2), \\
                \cdots \\
               \dot{o}_r=k_0(o_{r-1}o_r-o_ro_1),
    \end{cases}
\end{equation}
where $O_i$ denotes the clock signal species. For two networks, there needs to produce four oscillating signals, as displayed in Fig. \ref{fig_oscillator}, where the signals $O_1$ and $O_3$ serve as catalysts to participate reactions in (\ref{exp_CRC}) and (\ref{laycom_sigmoid}), respectively, 
 \begin{equation}\label{sigmoid_2_CRC}
        \begin{split}
        O_1+X + Y \overset{1}{\longrightarrow} O_1+X,\quad O_1+X \overset{1}{\longrightarrow} O_1;  \\  
         O_3+Y + Z \overset{1}{\longrightarrow} O_3+Y,\quad O_3+Z \ce{<=>[1][1]} O_3.
        \end{split}
    \end{equation}
As long as the phase length, i.e., the period $T$, satisfies that $T/2$ is greater than the reaction time needed for every network, the layer-by-layer dynamics can be obtained, and further works for computing the sigmoid function layer by layer.  

Although the introduction of oscillating signals renders a solution to compute the composite function layer by layer, it usually needs to produce many oscillating signals, especially in the case that the number of layers of composite function is large or a systematic learning task is implemented through molecular computation. For the latter case, our earlier work \cite{Fan_2025_automatic} indicated that there are $32$ oscillating signals needed to be constructed when implementing a three layer with two (input layer)-two (hidden layer)-one (output layer) nodes structured neural network. This will also add challenges for sophisticated molecular computation tasks, since it is rarely observed in natural biological systems to possess $32$ oscillating signals. 

To avoid the above drawbacks, we need to find other solutions to implement molecular computation of composite functions. On the one hand, the pattern of layer-by-layer computation should be kept; on the other hand, not too many oscillating signals should be inserted as catalysts.  

\subsection{ISS-based msCRCs composition}
The layer-by-layer computation provides a more general pattern on how to design CRNs for molecular computation. Essentially, this process relies on the composition rule of the composite function. A natural idea is to copy the composition rule of function to the composition of CRNs. From the viewpoint of msCRC, as an example of the composition of two msCRCs, it is requested that the output of the first msCRC should be the input of the second msCRC. We also utilize the notions given in \cite{Chalk_2019_composable}, and present ours from the viewpoint of dynamics.    

\begin{definition}[msCRCs Composition]\label{composition}
    Given $\mathscr{C}^\mathtt{1}=(\mathcal{S}^\mathtt{1},\mathcal{C}^\mathtt{1},\mathcal{R}^\mathtt{1},\kappa^\mathtt{1},\mathcal{X}^\mathtt{1},\mathcal{Y}^\mathtt{1})$ and $\mathscr{C}^\mathtt{2}=(\mathcal{S}^{\mathtt{2}},\mathcal{C}^{\mathtt{2}},\mathcal{R}^{\mathtt{2}},\kappa^{\mathtt{2}},\mathcal{X}^{\mathtt{2}},\mathcal{Y}^{\mathtt{2}})$ satisfying $\mathcal{Y}^\mathtt{1} \subseteq \mathcal{X}^{\mathtt{2}}$, the composition of $\mathscr{C}^\mathtt{1}$ and $\mathscr{C}^{\mathtt{2}}$ is the msCRC $\mathscr{C}^{\mathtt{2}\circ \mathtt{1}}=(\mathcal{S}^\mathtt{1} \cup \mathcal{S}^{\mathtt{2}}, \mathcal{C}^\mathtt{1} \cup \mathcal{C}^{\mathtt{2}}, \mathcal{R}^\mathtt{1} \cup \mathcal{R}^{\mathtt{2}}, \kappa^\mathtt{1} \cup \kappa^{\mathtt{2}},\mathcal{X}^\mathtt{1} \cup \mathcal{X}^{\mathtt{2}}, \mathcal{Y}^{\mathtt{2}})$. For the composition of more than three msCRCs, $\mathscr{C}^{p\circ(p-1)\circ \cdots \circ \mathtt{1}}$ is defined by the composition of $\mathscr{C}^{(p-1)\circ \cdots \circ \mathtt{1}}$ and $\mathscr{C}^p$ ($p\geq 3$).
\end{definition}

\begin{definition}[Dynamically Composable]\label{Def_dynamic_composable}
Assume $\mathscr{C}^\mathtt{1}=(\mathcal{S}^\mathtt{1},\mathcal{C}^\mathtt{1},\mathcal{R}^\mathtt{1},\kappa^\mathtt{1},\mathcal{X}^\mathtt{1},\mathcal{Y}^\mathtt{1})$ and $\mathscr{C}^\mathtt{2}=(\mathcal{S}^{\mathtt{2}},\mathcal{C}^{\mathtt{2}},\mathcal{R}^{\mathtt{2}},\kappa^{\mathtt{2}},\mathcal{X}^{\mathtt{2}},\mathcal{Y}^{\mathtt{2}})$ with $\mathcal{Y}^\mathtt{1}=\mathcal{X}^{\mathtt{2}},~\mathcal{Y}^{\mathtt{2}}\cap \mathcal{X}^{\mathtt{1}}=\varnothing$ and respective dynamics to be
\begin{equation}\label{dx=f}
\begin{cases}
  \dot{x}^\mathtt{1}=f^\mathtt{1}(x^\mathtt{1},y^\mathtt{1}), \\
  \dot{y}^\mathtt{1}=g^\mathtt{1}(x^\mathtt{1},y^\mathtt{1}),
\end{cases}x^\mathtt{1}(0)=x^\mathtt{1}_0,~y^\mathtt{1}(0)=y^\mathtt{1}_0
\end{equation} 
and
\begin{equation}\label{dy=g}
    \begin{cases}
    \dot{x}^\mathtt{2}=0, \\
    \dot{y}^\mathtt{2}=g^\mathtt{2}(x^\mathtt{2},y^\mathtt{2}),
    \end{cases} x^\mathtt{2}(0)=\bar{y}^\mathtt{1},~y^\mathtt{2}(0)=y^\mathtt{2}_0,
\end{equation}
and moreover, the system (\ref{dx=f}) supports $\lim_{t \to \infty} y^\mathtt{1}(t)= \bar{y}^\mathtt{1}$, and the system (\ref{dy=g}) satisfies $\lim_{t \to \infty} y^{\mathtt{2}}(t)= \bar{y}^{\mathtt{2}}$. If the solution (output components) of the coupled system 
\begin{equation}\label{coupled_system}
    \begin{cases}
     \dot{x}^\mathtt{1}=f^\mathtt{1}(x^\mathtt{1},y^\mathtt{1}), \\
     \dot{y}^\mathtt{1}=g^\mathtt{1}(x^\mathtt{1},y^\mathtt{1}),\\
   \dot{y}^\mathtt{2}=g^\mathtt{2}(x^\mathtt{2},y^\mathtt{2}), 
    \end{cases} \begin{array}{l}
         x^\mathtt{1}(0)=x^\mathtt{1}_0,~y^\mathtt{1}(0)=y^\mathtt{1}_0, \\
         y^\mathtt{2}(0)=y^\mathtt{2}_0,
    \end{array}
\end{equation}
satisfies 
\begin{equation}\label{dynamic_composable}
    \lim_{t \rightarrow \infty} \left [ \begin{array}{c}
         y^\mathtt{1}(t)  \\
         y^\mathtt{2}(t)  \\
    \end{array} \right] =\left[ \begin{array}{c}
         \bar{y}^\mathtt{1} \\
         \bar{y}^\mathtt{2} \\
    \end{array} \right],
\end{equation}
then the two msCRCs are said to be dynamically composable. We say $p~(p\ge3)$ msCRCs $\mathscr{C}^{\mathtt{i}}~(1\le \mathtt{i} \le p)$ are dynamically composable, if for any $2\le \mathtt{j} \le p$, $\mathscr{C}^{(\mathtt{j-1})\circ \cdots \circ 1}$ and $\mathscr{C}^{\mathtt{j}}$ are dynamically composable.
\end{definition}

\begin{remark}
 The above definition indicates that for two dynamically composable msCRCs (\ref{dx=f}) and (\ref{dy=g}), the output limiting steady state of their coupled system (\ref{coupled_system}) can be thought as a layer-by-layer composition of their individual output limiting steady state. This observation is very helpful for the implementation of layer-by-layer molecular computations. 
\end{remark}    

\begin{theorem}\label{main_thm}
    Let msCRCs $\mathscr{C}^\mathtt{1}=(\mathcal{S}^\mathtt{1},\mathcal{C}^\mathtt{1},\mathcal{R}^\mathtt{1},\kappa^\mathtt{1},\mathcal{X}^\mathtt{1},\mathcal{Y}^\mathtt{1})$ and $\mathscr{C}^\mathtt{2}=(\mathcal{S}^{\mathtt{2}},\mathcal{C}^{\mathtt{2}},\mathcal{R}^{\mathtt{2}},\kappa^{\mathtt{2}},\mathcal{X}^{\mathtt{2}},\mathcal{Y}^{\mathtt{2}})$ be dynamic computations of funtions $\sigma_1:\mathbb{R}^{m_1}_{\ge 0} \to \mathbb{R}^{n_1-m_1}_{\ge 0}$ and $\sigma_2:\mathbb{R}^{m_2}_{\ge 0} \to \mathbb{R}^{n_2-m_2}_{\ge 0}$, respectively. Then their composition $\mathscr{C}^{\mathtt{2}\circ \mathtt{1}}$ is a dynamic computation of $\sigma_2 \circ \sigma_1:\mathbb{R}^{m_1}_{\ge 0} \to \mathbb{R}^{n_2-m_2}_{\ge 0}$ if $\mathscr{C}^\mathtt{1}$ and $\mathscr{C}^\mathtt{2}$ are dynamically composable.
\end{theorem}

The detailed proof appears in Appendix \ref{appendixA}, which may apply to all propositions, theorems and corollaries. The result can be easily extended to suit for more msCRCs.
\begin{corollary}\label{coro_thm1}
    Let msCRCs $\mathscr{C}^\mathtt{i}=(\mathcal{S}^\mathtt{i},\mathcal{C}^\mathtt{i},\mathcal{R}^\mathtt{i},\kappa^\mathtt{i},\mathcal{X}^\mathtt{i},\mathcal{Y}^\mathtt{i})$ be dynamic computations of functions $\sigma_i:\mathbb{R}^{m_i}_{\ge 0} \to \mathbb{R}^{n_i-m_i}_{\ge 0}$, respectively, where $1\le \mathtt{i} \le p$. Then their composition $\mathscr{C}^{\mathtt{p}\circ \cdots \circ \mathtt{2}\circ \mathtt{1}}$ is a dynamic computation of $\sigma_p \circ \sigma_{p-1} \circ \cdots \circ \sigma_1:\mathbb{R}^{m_1}_{\ge 0} \to \mathbb{R}^{n_p-m_p}_{\ge 0}$ if for any $2\le \mathtt{j} \le p$, $\mathscr{C}^{(\mathtt{j-1})\circ \cdots \circ 1}$ and $\mathscr{C}^{\mathtt{j}}$ are dynamically composable.
\end{corollary}

\textit{Theorem} \ref{main_thm} and \textit{Corollary} \ref{coro_thm1} clarify that the notion of dynamic composability plays an important role on performing layer-by-layer molecular computations for a composite function. This encourages us to find the conditions to induce dynamic composability for two msCRCs.

\begin{proposition}\label{prop_simple_structure}
    Given $\mathscr{C}^\mathtt{1}=(\mathcal{S}^\mathtt{1},\mathcal{C}^\mathtt{1},\mathcal{R}^\mathtt{1},\kappa^\mathtt{1},\mathcal{X}^\mathtt{1},\mathcal{Y}^\mathtt{1})$ and $\mathscr{C}=(\mathcal{S},\mathcal{C},\mathcal{R},\kappa,\mathcal{X},\mathcal{Y})$ satisfying $\mathcal{Y}^\mathtt{1}=\mathcal{X},~\mathcal{Y}\cap \mathcal{X}^{\mathtt{1}}=\varnothing$, if for $\forall X\in\mathcal{X},~\forall Y\in \mathcal{Y}$, their dynamics has the form of
        \begin{equation}\label{dy=p-qy}
            \begin{cases}
            \dot{x}=0, \\
            \dot{y}= p(x)-q(x)y, 
            \end{cases}
        \end{equation}
    where $x \in \mathbb{R}_{\ge 0}^m$, $p: \mathbb{R}_{\ge 0}^m \to \mathbb{R}_{\ge 0},~q:\mathbb{R}_{\ge 0}^m \to \mathbb{R}_{> 0}$, then $\mathscr{C}^{\mathtt{1}}$ and $\mathscr{C}$ are dynamically composable.
\end{proposition}

The dynamical equation (\ref{dy=p-qy}) expresses a linear relation on the concentration of each output species, which also implies that the stoichiometric coefficient of $Y$ as a reactant in this msCRC should be $0$ or $1$. We use this proposition to check the msCRC (\ref{laycom_sigmoid}) in \textit{Example} \ref{ex_sigmoid}. From $\dot{z}=1-(1+y)z$, we confirm that the msCRCs (\ref{exp_CRC}) and (\ref{laycom_sigmoid}) are dynamically composable. Further, based on \textit{Theorem} \ref{main_thm}, they can be used to compute the sigmoid function layer by layer, i.e., firstly computing $y=e^{-x}$, and then computing $z=\frac{1}{1+y}$.

Note that essentially the three systems (\ref{dx=f}), (\ref{dy=g}) and (\ref{coupled_system}) related to dynamic composability are all MASs, so the state of each system should evolve in the nonnegative stoichiometric compatibility class of their corresponding initial value. It is straightforward to get $\bar{y}^\mathtt{1}-y_0^\mathtt{1}\in \mathscr{S}_{\mathtt{1}}^{y^\mathtt{1}}$, $(\bar{y}^\mathtt{1},\bar{y}^\mathtt{2})^\top-(y_0^\mathtt{1},y_0^\mathtt{2})^\top\in \mathscr{S}_{\mathtt{c}}^{y^\mathtt{1},y^\mathtt{2}}$, which indicates that the output concentrations of the systems (\ref{dx=f}) and (\ref{coupled_system}) are related to a certain kind of stability. Here, $\mathscr{S}_a^b$ is the truncation of $\mathscr{S}_a$ along $b$. We can also derive easily that $\bar{y}^\mathtt{2}-y_0^\mathtt{2}\in \mathscr{S}_{\mathtt{2}}^{y^\mathtt{2}}$, which likewise means the output in (\ref{dy=g}) evolves in a kind of stable pattern. All of these impressions suggest that the three systems related to dynamic composability of msCRCs should exhibit stable behaviors in the output components. Especially, for the msCRC $\mathscr{C}^\mathtt{2}$ ruled by (\ref{dy=g}), the stable output relies on the condition of input $x=\bar{x}$, which motivates us to consider the notion of ISS to characterize dynamic composability. 

\begin{theorem}\label{Thm_ISS_net}
Let $\mathscr{C}^\mathtt{1}=(\mathcal{S}^\mathtt{1},\mathcal{C}^\mathtt{1},\mathcal{R}^\mathtt{1},\kappa^\mathtt{1},\mathcal{X}^\mathtt{1},\mathcal{Y}^\mathtt{1})$ and $\mathscr{C}^\mathtt{2}=(\mathcal{S}^{\mathtt{2}},\mathcal{C}^{\mathtt{2}},\mathcal{R}^{\mathtt{2}},\kappa^{\mathtt{2}},\mathcal{X}^{\mathtt{2}},\mathcal{Y}^{\mathtt{2}})$, ruled by (\ref{dx=f}) and (\ref{dy=g}), respectively, satisfy $\mathcal{Y}^\mathtt{1}=\mathcal{X}^{\mathtt{2}},~\mathcal{Y}^{\mathtt{2}}\cap \mathcal{X}^{\mathtt{1}}=\varnothing$ and $\lim_{t \to \infty} y^\mathtt{1}(t)= \bar{y}^\mathtt{1}$, $\lim_{t \to \infty} y^{\mathtt{2}}(t)= \bar{y}^{\mathtt{2}}$. Consider the $y^\mathtt{2}$-related part of the system (\ref{dy=g}) independently, i.e., 
\begin{equation}\label{g2_y}
  \dot{y}^\mathtt{2}=g^\mathtt{2}(x^\mathtt{2},y^\mathtt{2}).  
\end{equation}
If it is ISS regarding $(\bar{y}^{\mathtt{1}},\bar{y}^{\mathtt{2}})$, that is, there exist a class $\mathcal{KL}$ function $\beta:\mathbb{R}_{\ge 0} \times \mathbb{R}_{\ge 0} \to \mathbb{R}_{\ge 0}$ and a class $\mathcal{K}$ function $\gamma:\mathbb{R}_{\ge 0} \to \mathbb{R}_{\ge 0}$ such that for $\forall y^{\mathtt{2}}_0$ and any bounded input $x^{\mathtt{2}}(t)$, the solution $y^{\mathtt{2}}(t)$ exists for all $t \ge 0$ and satisfies
\begin{equation*}
   \left \| y^{\mathtt{2}} \left( t \right) -\bar{y}^{\mathtt{2}} \right \| \le \beta \left( \left\| y^{\mathtt{2}}_0 - \bar{y}^{\mathtt{2}} \right\| , t \right) + \gamma ( \sup \limits_{0\le \tau \le t}\left\| x^{\mathtt{2}} \left( \tau \right) -\bar{y}^{\mathtt{1}} \right\| ),
\end{equation*}   
then $\mathscr{C}^\mathtt{1}$ and $\mathscr{C}^\mathtt{2}$ are dynamically composable.
\end{theorem}

Similarly, we can get the corresponding result in the case of more than three msCRCs. 
\begin{corollary}\label{coro_thmISS}
    Let $\mathscr{C}^\mathtt{1}=(\mathcal{S}^\mathtt{1},\mathcal{C}^\mathtt{1},\mathcal{R}^\mathtt{1},\kappa^\mathtt{1},\mathcal{X}^\mathtt{1},\mathcal{Y}^\mathtt{1})$ be ruled by (\ref{dx=f}), and $\mathscr{C}^\mathtt{i}=(\mathcal{S}^\mathtt{i},\mathcal{C}^\mathtt{i},\mathcal{R}^\mathtt{i},\kappa^\mathtt{i},\mathcal{X}^\mathtt{i},\mathcal{Y}^\mathtt{i}), ~\mathtt{i}=2,...,p$, be ruled by 
    \begin{equation}\label{dyi=g}
        \begin{cases}
        \dot{x}^\mathtt{i}=0, \\
        \dot{y}^\mathtt{i}=g^\mathtt{i}(x^\mathtt{i},y^\mathtt{i}),
        \end{cases} x^\mathtt{i}(0)=\bar{y}^\mathtt{i-1},~y^\mathtt{i}(0)=y^\mathtt{i}_0.
    \end{equation}
   Suppose they satisfy
    \begin{enumerate}
        \item $\mathcal{Y}^{\mathtt{i-1}}=\mathcal{X}^{\mathtt{i}},~2\le \mathtt{i} \le p$;
        \item $\left(\bigcup_{\mathtt{j}=1}^{i-1}\mathcal{X}^{\mathtt{j}}\right) \cap \mathcal{Y}^{\mathtt{i}}=\varnothing,~2\le \mathtt{i} \le p$;
        \item $\lim_{t \to \infty}y^{\mathtt{i}}(t)=\bar{y}^{\mathtt{i}},~1 \le \mathtt{i} \le p$.
    \end{enumerate}
    If the $y^{\mathtt{i}}$-related part of the system (\ref{dyi=g}) is ISS regarding $(\bar{y}^{\mathtt{i-1}},\bar{y}^{\mathtt{i}}), ~2 \le \mathtt{i} \le p$, then for any $2\le \mathtt{j} \le p$, $\mathscr{C}^{(\mathtt{j-1})\circ \cdots \circ 1}$ and $\mathscr{C}^{\mathtt{j}}$ are dynamically composable.
\end{corollary}

Naturally, the conditions to suggest ISS \cite{Khalil_2002_nonlinear} can enrich the above results. 
\begin{proposition}\label{iss_condition}
Consider the same two msCRCs $\mathscr{C}^\mathtt{1},~\mathscr{C}^\mathtt{2}$ as given in \textit{Theorem} \ref{Thm_ISS_net}. Then $\mathscr{C}^\mathtt{1}$ and $\mathscr{C}^\mathtt{2}$ are dynamically composable if one of the following conditions is true
\begin{itemize}
    \item [1)] $g^\mathtt{2}(x^\mathtt{2},y^\mathtt{2})$ in (\ref{g2_y}) is continuously differentiable and globally Lipschitz in $(x^\mathtt{2},y^\mathtt{2})$, and $\bar{y}^{\mathtt{2}}$ is a globally exponentially stable equilibrium point in $\mathscr{C}^\mathtt{2}$.
    \item [2)] the system  (\ref{g2_y}) admits an ISS-Lyapunov function $V: \mathbb{R}^{n_2-m_2} \to \mathbb{R}_{\ge 0}$, which is smooth, positive definite, and radially unbounded, and also satisfies that
    \begin{itemize}
        \item [] for bounded input $x^{\mathtt{2}}$, there exist a positive definite function $W:\mathbb{R}^{n_2-m_2} \to \mathbb{R}_{\ge 0}$ and a class $\mathcal{K}$ function $\rho:\mathbb{R}_{\ge 0} \to \mathbb{R}_{\ge 0}$ to support
         \begin{equation}
            \nabla V^{\top} (y^{\mathtt{2}}-\bar{y}^{\mathtt{2}})g^{\mathtt{2}}(x^{\mathtt{2}},y^{\mathtt{2}}) \le -W(y^{\mathtt{2}}-\bar{y}^{\mathtt{2}})
    \end{equation}
    under the condition of $\left \|y^{\mathtt{2}}-\bar{y}^{\mathtt{2}} \right  \| \ge \rho (\left\|x^{\mathtt{2}}-\bar{x}^{\mathtt{2}} \right \|)$.
    \end{itemize}   
\end{itemize}
\end{proposition}

It should be noted that the first condition of \textit{Proposition} \ref{iss_condition} can seldom be true for MASs, but it may remain applicable for analyzing enzymatic reaction systems characterized by fractional-structured Michaelis-Menten kinetics. We use the following example to exhibit the application of \textit{Theorem} \ref{Thm_ISS_net} and \textit{Proposition} \ref{iss_condition}. 
\begin{example}\label{ex_sqrt}
    Consider the following two msCRCs
    \begin{equation}\label{sqrt_coupled_system}
    \begin{split}
        \mathscr{C}^{\mathtt{1}}:&~Y \ce{<=>[1][1]} \varnothing,   ~~~~~~~~X \overset{1}{\longrightarrow} X+Y, \\
        \mathscr{C}^\mathtt{2}:&~3Z \overset{1}{\longrightarrow} 2Z, ~~~~~~~  Y \overset{1}{\longrightarrow} Y+Z, \\
    \end{split}
    \end{equation}
and set $\mathcal{X}^\mathtt{1}=\{X\},~\mathcal{Y}^\mathtt{1}=\mathcal{X}^\mathtt{2}=\{Y\},~\mathcal{Y}^\mathtt{2}=\{Z\}$. The coupled dynamics takes
    \begin{equation*}
        \begin{cases}
            \dot{x}=0, \\
            \dot{y}=x+1-y, \\
            \dot{z}=y-z^3,
        \end{cases}\begin{array}{l}
        x(0)=x_0,~y(0)=y_0, \\
         z(0)=z_0.
    \end{array}
    \end{equation*}
Clearly, $\mathscr{C}^\mathtt{1}$ and $\mathscr{C}^\mathtt{2}$ are dynamic computations of $y=x+1$ and $z=\sqrt[3]{y}$, respectively, and $\bar{y}=x_0+1,~\bar{z}=\sqrt[3]{x_0+1}$. 

Next, we check if the system $\dot{z}=y-z^3$ is ISS regarding $(\bar{y},\bar{z})=(x_0+1,\sqrt[3]{x_0+1})$. Construct a function $$V(z-\bar{z})=\frac{1}{2}(z-\bar{z})^2,$$ which is obviously smooth, positive definite, and radially unbounded. Then we have
 \begin{equation}\label{dv/dt}
    \begin{split}
        &\frac{\mathrm{d}V}{\mathrm{d}t} =(z-\bar{z})\left(y-z^3 \right) \\
               =&-(z-\bar{z})^2\left[(z-\bar{z})^2+3\bar{z}(z-\bar{z})+3\bar{z}^2\right] \\
         &+(z-\bar{z})(y-\bar{y})  \\
        \le &-\frac{3}{4}\bar{z}^2(z-\bar{z})^2+|z-\bar{z}||y-\bar{y}|\\ 
        = & -\alpha V+|z-\bar{z}||y-\bar{y}|,
    \end{split}
    \end{equation}
    where $\alpha=\frac{3}{2}\bar{z}^2>0$. From the network $\mathscr{C}^2$, it is straightforward that $z(t)$ is uniformly bounded in time when $y(t)$ is bounded (suggested by $\mathscr{C}^1$), which implies $\exists~C>0$ such that $|z-\bar{z}|\le C$. We can thus get $\frac{\mathrm{d}V}{\mathrm{d}t}\le -\alpha V+C|y-\bar{y}|.$ Further, we have
   $\frac{\mathrm{d}}{\mathrm{d}t}\left(e^{\alpha t}V\right) \le Ce^{\alpha t}|y-\bar{y}|.$ By integration, we obtain
   $ V(t)\le \left(V(0)+C\int_0^te^{\alpha s}|y(s)-\bar{y}|\mathrm{d}s\right)e^{-\alpha t}.$ This results in
     \begin{align*}
        |z(t)-\bar{z}|  & \le \sqrt{2\left(V(0)+C\int_0^te^{\alpha s}|y(s)-\bar{y}|\mathrm{d}s\right)e^{-\alpha t}} \\
       & \le \sqrt{2V(0)e^{-\alpha t}} + \sqrt{2Ce^{-\alpha t}\int_0^te^{\alpha s}|y(s)-\bar{y}|\mathrm{d}s} \\
         &\le |z(0)-\bar{z}|e^{-\frac{\alpha}{2}t}+\sqrt{CMe^{-\alpha t}\int_0^te^{\alpha s}\mathrm{d}s} \\
        &\le |z(0)-\bar{z}|e^{-\frac{\alpha}{2}t}+\sqrt{\frac{CM}{\alpha}} \\
        &=\beta(|z(0)-\bar{z}|,t)+\gamma\left(\sup_{0 \le s \le t}|y(s)-\bar{y}|\right),
    \end{align*}
   where $M=2\sup_{0 \le s \le t}|y(s)-\bar{y}|$, $\beta(s,t)=se^{-\frac{\alpha}{2}t}$ is a class $\mathcal{KL}$ function and $\gamma(s)=\sqrt{\frac{2C}{\alpha}s}$ is a class $\mathcal{K}$ function. We thus conclude that $\dot{z}=y-z^3$ is ISS regarding $(\bar{y},\bar{z})$, and by \textit{Theorem} \ref{Thm_ISS_net}, $\mathscr{C}^\mathtt{1}$ and $\mathscr{C}^\mathtt{2}$ are dynamically composable. They serve to dynamically compute the function $z=\sqrt[3]{x+1}$.
\end{example}

The example to demonstrate the computation of composite function through more than two msCRCs is presented in Appendix \ref{appendixB}.

\section{Comparisons with state-of-the-art methods}\label{Section_4}
In this section, we compare our method with rate-independent CRNs composition \cite{Chalk_2019_composable} and oscillating signal-catalyzed CRNs composition \cite{Fan_2024_automatic} to demonstrate their differences. Compared to the former, our method can handle the computation of more sophisticated composite functions, while in comparison to the latter, our method exhibits to be able to avoid introducing more species, and also provide a more general computation pattern according to the composition rules of functions.  
\subsection{Comparison with composable rate-independent CRNs}
Rate-independent CRNs composition provides an intuitive way to compute some composite functions. The main rule to evaluate the computable function is the following necessary and sufficient condition. 
\begin{theorem}[Chalk \textit{et al.} \cite{Chalk_2019_composable}]\label{thm_rate_independent}
    A function $f:\mathbb{R}^n_{\ge 0} \to \mathbb{R}_{\ge 0}$ is computable by a composable chemical reaction computer if and only if $f$ is superadditive positive-continuous piecewise rational linear.
\end{theorem}

\begin{figure}
    \centering
    \includegraphics[width=0.47\textwidth]{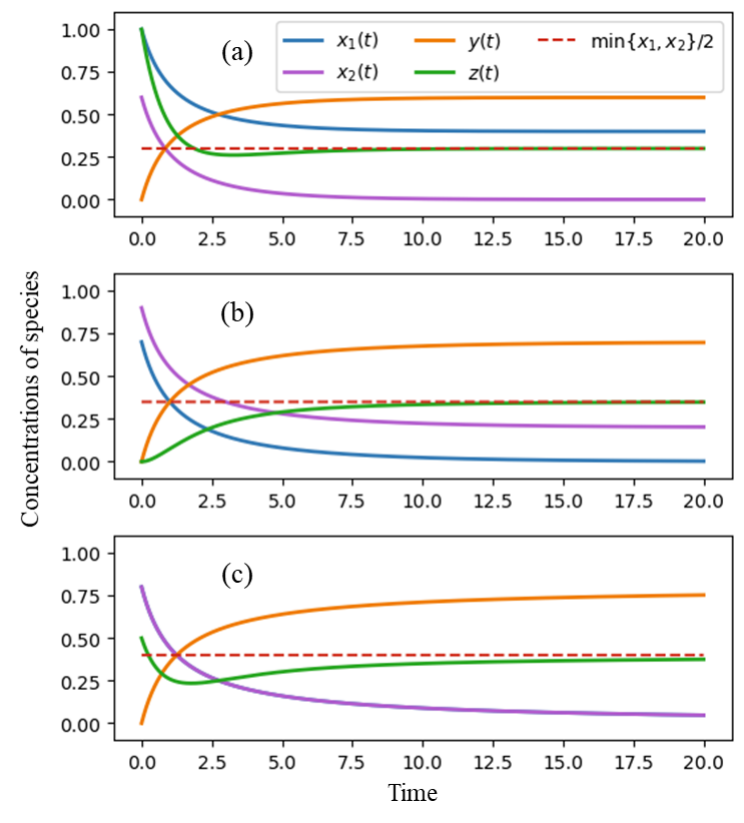}
    \caption{The evolution of the concentration of each species in (\ref{minFun}) with three kinds of different initial values: (a) $x_1(0)=1.0>x_2(0)=0.6,~y(0)=0,~z(0)=1.0$; (b) $x_1(0)=0.7<x_2(0)=0.9,~y(0)=0,~z(0)=0$; (c) $x_1(0)=x_2(0)=0.8,~y(0)=0,~z(0)=0.5$ (Under this condition, the curves of $x_1(t)$ and $x_2(t)$ overlap).}
    \label{fig:minFun}
\end{figure}

Essentially, this method can only handle some simple functions, and moreover, they should be constrained by the mentioned conditions in the theorem, such as the minimum-solving related function.  
\begin{example}
For the function
    \begin{equation}\label{rational_linear_fun}
        f(x_1,x_2)=\frac{1}{2}\min\{x_1,x_2\}, 
    \end{equation}
\end{example}
which is superadditive positive-continuous piecewise rational linear, we can design the rate-independent CRN 
\begin{equation}
    X_1+X_2 \longrightarrow Y, \quad 2Y \longrightarrow Z
\end{equation}
to compute it. It is readily to infer that the limiting steady state of species $Z$, $\bar{z}=\frac{1}{2}\min\{x_{1}(0),x_{2}(0)\}$, stores the computation result. Naturally, this computation can also be implemented through our method. Construct the MASs as follows:  
\begin{align*}
    \mathscr{C}^{\mathtt{1}}:~X_1+X_2 \overset{1}{\longrightarrow}Y;\quad
    \mathscr{C}^{\mathtt{2}}:~Y \overset{\frac{1}{2}}{\longrightarrow}Y+Z, \quad Z \overset{1}{\longrightarrow} \varnothing.
\end{align*}
The coupled dynamics is
\begin{equation}\label{minFun}
    \begin{cases}
          \dot{x}_1=-x_1x_2,\quad \dot{x}_2=-x_1x_2,\\
       \dot{y}=x_1x_2, \quad \dot{z}=\frac{1}{2}y-z,
    \end{cases}
\end{equation}
which, based on \textit{Proposition} \ref{prop_simple_structure} and \textit{Theorem} \ref{main_thm}, finishes the computation task through suggesting $\bar{y}=\min\{x_{1}(0),x_{2}(0) \}$ and $\bar{z}=\frac{1}{2}\bar{y}$. Fig. \ref{fig:minFun} displays the evolution of the concentration of each species in (\ref{minFun}) under three kinds of different initial conditions $x_1(0),x_2(0)$. In all cases, $\bar{z}$ carries the final computation result $\frac{1}{2}\min\{x_{1}(0),x_{2}(0)\}$. Note that we use three different initial values of $z(0)$ to show the computation results to be independent of them. We solve the dynamical equations (\ref{minFun}) using the ode solver from the scipy.integrate module in Python 3.10.9, which also apply to other simulations in this paper.

The special request on the computed functions limits the application of the rate-independent CRNs composition method to more and/or more sophisticated composite functions. Even though for the sigmoid function $z=\frac{1}{1+e^{-x}}$ in \textit{Example} \ref{ex_sigmoid} and the function $z=\sqrt[3]{x+1}$ in \textit{Example} \ref{ex_sqrt}, this method is of helplessness. In contrast to our method, the utilization of dynamic composability provides a larger room to compute these sophisticated composite functions by msCRCs. The inherent reason is that, logically, any computation can be embedded into a class of polynomial ODEs, and then implement these ODEs with MASs \cite{Fages_2017_strong}. Therefore, the proposed msCRCs composition method has the potential to compute more sophisticated composite functions layer by layer according to the composition rule of functions.

\subsection{Comparison with oscillating signals-catalyzed CRNs}
\begin{figure}
    \centering
    \includegraphics[width=0.47\textwidth]{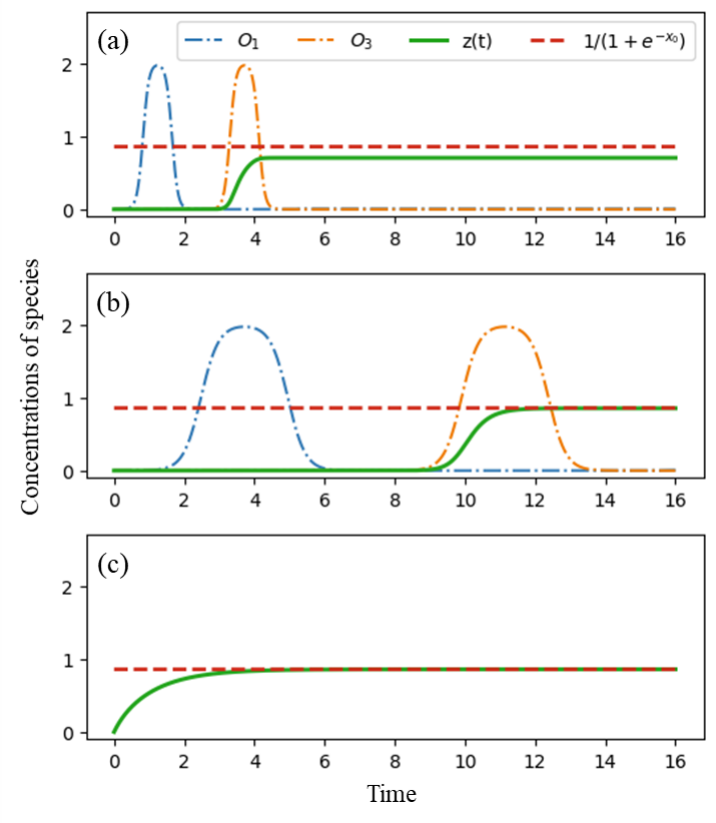}
    \caption{The computation results of the sigmoid function with $x_0=1.8,y_0=1,z_0=0$: (a) Oscillating signals-catalyzed CRNs method with $k_0=6$ corresponding to $T\approx 5.3$; (b) with $k_0=2$ corresponding to $T\approx 15.9$; (c) Our method.}
    \label{fig:oscill_sigmoid}
\end{figure}

For the aforementioned similar functions, except for our method, the oscillating signals-catalyzed CRNs is another feasible solution to implement their computations by molecular. As an example of the sigmoid function $z=\frac{1}{1+e^{-x}}$, the network in (\ref{sigmoid_2_CRC}) is able to perform the computation task, where two new species $O_1,~O_3$ are introduced as catalysts to participate reactions. Their oscillatory behaviors are utilized to switch reactions on and off, thereby controlling the occurrence sequence of reactions to achieve the goal of sequential computation. Indeed, this method can finish the layer-by-layer computation of the sigmoid function according to its composition rule as long as the phase length $T$ is twice as much as the time needed for each reaction in the network. Fig. \ref{fig:oscill_sigmoid} presents the related computation results (the curves of $z(t)$) with two different phase lengths. In sub-figure (a)/(b), $T\approx 5.3$/$T\approx 15.9$ that is calculated through setting $k_0=6$/$k_0=2$ in the MAS (\ref{oscillator}). As can be seen, if $T$ is not set properly, such as $T\approx 5.3$, there will lead to large error in the computation result. However, when the phase length is set $T\approx 15.6$, the computation result approaches the true one after $t\approx 12$. To make a comparison, Fig. \ref{fig:oscill_sigmoid}(c) also provides the computation result through our method. Apparently, $z(t)$ asymptotically changes to $\bar{z}$ from $z_0=0$, and there is no horizontal interval during the beginning stage. This means that the duration for species $Z$ to participates reaction occupies the total reaction time, and its concentration can approach the computation result soon from the beginning.   

The above observation indicates that, for the oscillating signals-catalyzed CRNs method, the main challenges lie in the difficulty to produce enough oscillating signals in practical biological systems and the inducing error from using $z(T)$ to approximate $\bar{z}=\lim_{t\to \infty}z(t)$. For the latter, since some reactions have to be shut down at $t=T$, we need to evaluate $\bar{z}$ at a finite time, i.e., $\bar{z}\approx z(T)$. Moreover, this kind of approximation errors will further accumulate and/or propagate along computation procedures, which is specially true for implementing complicated computation tasks \cite{Fan_2024_automatic}. Naturally, the approximation errors are also present in our method. But the evaluation time to approximate $\bar{z}$ is not fixed at a certain time, like $T$ in the oscillating signals-catalyzed CRNs method. The following theorem quantitatively characterizes the error estimation between these two kinds of methods when they are applied to compute the sigmoid function with $z(T)$ as an estimation.

\begin{theorem}\label{thm_error_n2}
   Consider the computation task of the sigmoid function $z=\frac{1}{1+e^{-x}}$. Suppose $T>0$ is the finite period of the chemical oscillator (\ref{oscillator}), then the error estimation of the computation result with the oscillator method, i.e., using the msCRC (\ref{sigmoid_2_CRC}), and the dynamic composition method, i.e., using the msCRCs (\ref{exp_CRC}) and (\ref{laycom_sigmoid}), individually satisfies
   \begin{enumerate}
    \item for the oscillator method, $\exists ~C_1>0$ such that
    \begin{equation}\label{2_step_esti}
        \left| z\left( T \right) -\frac{1}{1+e^{-x_0}} \right| \le C_1e^{-\frac{T}{2}};
    \end{equation}
    \item for our method, $\forall ~\varepsilon >0$, $\exists ~C_2,C_3>0$ such that
    \begin{equation}\label{1_step_esti}
        \left| z\left( T \right) -\frac{1}{1+e^{-x_0}} \right| \le C_2e^{-(1-\varepsilon)T}+C_3e^{-T}.
    \end{equation}
    \end{enumerate}
\end{theorem}

\begin{figure}[htbp]
      \centering
      \includegraphics[width=0.48\textwidth]{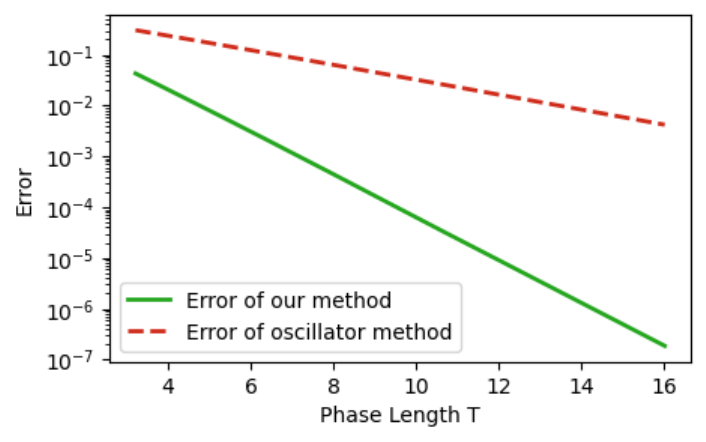}
      \caption{Errors comparison of computing sigmoid function between the oscillator method and our method with $x_0=1.8,y_0=1,z_0=0$.}
      \label{fig:error}
\end{figure}

 \textit{Theorem} \ref{thm_error_n2} shows that our method has a faster convergence rate than the oscillator method, especially when the phase length is small. Certainly, their difference will reduce as $T$ increases. To observe this difference visually, we provide the error change along the phase length in Fig. \ref{fig:error} for these two methods. The curve trends validate the error estimation presented in \textit{Theorem} \ref{thm_error_n2}.

\section{Conclusions}\label{Section_5}
In this paper, we aim at following the composition rule of functions to design CRNs to implement molecular computations. By borrowing the concepts of msCRC and rate-independent CRNs composition \cite{Chalk_2019_composable}, we define the notions of msCRC, dynamic computation and dynamic composability. Based on them, we realize the layer-by-layer molecular computations for some composite functions accompanied by theoretical supports to suggest two msCRCs (or more) able to implement composition computations. Further, we utilize the notion of ISS, and propose several sufficient conditions to say two or more msCRCs dynamically composable. These allow us to identify what msCRCs can be combined to finish the desired molecular computations. The proposed method exhibits some advantages in computation ability and computation accuracy, compared to the state-of-the-art methods. We use several examples to illustrate this point.

Despite the above achievements, there is still a large room to strengthen the current dynamically composable msCRCs method for molecular computations. Intuitively, the application of ISS and its related control theory to this subject is superficial, which leaves a door to explore more ISS-evaluated conditions for possible applications. Another important point is to link the network structure to the ISS property, which may open a larger door to design dynamically composable msCRCs with special topological structures. In addition, since many composite functions can be generated through compounding a few fundamental elementary functions, future research could explore the composability of msCRCs corresponding to fundamental elementary functions. It will be quite valuable to build a ``composable fundamental elementary msCRC library".

\appendices
\section*{Appendix}
In this appendix, we provide the proofs for the theorems, corollaries and propositions presented previously in part A and give an example to compute a composite function through three msCRCs in part B. 
\setcounter{equation}{0}
\renewcommand\theequation{A.\arabic{equation}}

\subsection{Proofs}\label{appendixA}

\noindent \textbf{Proof of Theorem \ref{main_thm}: }

\begin{proof}
    Let $x^{\mathtt{i}}, y^{\mathtt{i}}$ be the concentration vectors of species in $\mathcal{X}^{\mathtt{i}},\mathcal{Y}^{\mathtt{i}}$, respectively, where $\mathtt{i}=1,2$. Then the dynamics of $\mathscr{C}^{\mathtt{i}}$ is given by
    $$\begin{cases}
        \dot{x}^{\mathtt{i}}=f^{\mathtt{i}}(x^{\mathtt{i}},y^{\mathtt{i}}), \\
        \dot{y}^{\mathtt{i}}=g^{\mathtt{i}}(x^{\mathtt{i}},y^{\mathtt{i}}), 
    \end{cases} x^{\mathtt{i}}(0)=x_0^{\mathtt{i}},~y^{\mathtt{i}}(0)=y_0^{\mathtt{i}},$$ 
    and its solutions satisfy $\bar{y}^\mathtt{i}=\lim_{t \to \infty} y^\mathtt{i}(t)=\sigma_i(x_0^{\mathtt{i}})$, where $\mathtt{i}=1,2$. Since $\mathscr{C}^{\mathtt{1}}$ and $\mathscr{C}^{\mathtt{2}}$ are dynamically composable, it takes $f^\mathtt{2}=0$, and the dynamics of $\mathscr{C}_{2 \circ 1}$ is given by
    $$\begin{cases}
        \dot{x}^{\mathtt{1}}=f^{\mathtt{1}}(x^{\mathtt{1}},y^{\mathtt{1}}),  \\
        \dot{y}^{\mathtt{1}}=g^{\mathtt{1}}(x^{\mathtt{1}},y^{\mathtt{1}}),  \\
        \dot{y}^{\mathtt{2}}=g^{\mathtt{2}}(y^{\mathtt{1}},y^\mathtt{2}), 
    \end{cases} \begin{array}{l}
         x^{\mathtt{1}}(0)=x_0^{\mathtt{1}},~y^{\mathtt{1}}(0)=y_0^{\mathtt{1}}, \\
         y^{\mathtt{2}}(0)=y_0^{\mathtt{2}}.
    \end{array}$$
    Due to Definition \ref{Def_dynamic_composable}, it holds 
    $$\lim_{t\to \infty}y^{\mathtt{2}}(t)=\sigma_2(\bar{y}^{\mathtt{1}})=\sigma_2(\sigma_1(x_0^{\mathtt{1}})),$$
    which means $\mathscr{C}_{2\circ 1}$ is a dynamic computation of $\sigma_2 \circ \sigma_1$.
\end{proof}

\noindent \textbf{Proof of Corollary \ref{coro_thm1}: }

\begin{proof}
    The proof proceeds by induction on $p$. The result for $p=2$ is described as \textit{Theorem} \ref{main_thm} before. Suppose that the result holds for $p=r (r \ge 2)$. Then the dynamics of $\mathscr{C}_{r\circ \cdots \circ 1}$ is given by
    $$\begin{cases}
        \dot{x}=f(x,y^{\mathtt{r}}),  \\
        \dot{y}^{\mathtt{r}}=g^{\mathtt{r}}(x,y^{\mathtt{r}}), & 
    \end{cases},x(0)=x_0,~y^{\mathtt{r}}(0)=y_0^{\mathtt{r}}.$$
    where $x$ denotes the concentrations of all the species in $\mathscr{C}_{r\circ \cdots \circ 1}$, and its solutions satisfy $\bar{y}^{\mathtt{r}}=\lim_{t \to \infty} y^{\mathtt{r}}(t)=(\sigma_{r} \circ \cdots \circ \sigma_1)(x_0^{\mathtt{1}})$. Applying Theorem \ref{main_thm} by taking $\mathscr{C}_{r\circ \cdots \circ 1}$ as the first msCRC and $\mathscr{C}_{r+1}$ as the second msCRC, we can conclude that $\mathscr{C}_{(r+1)\circ r \circ \cdots \circ 1}$ is a dynamic computation of $\sigma_{r+1} \circ \sigma_r \circ \cdots \circ \sigma_1$, and the proof is complete.
\end{proof}

\noindent \textbf{Proof of Proposition \ref{prop_simple_structure}: }

\begin{proof}
    For system (\ref{dy=p-qy}) with initial values $x(0)=\bar{x},~y(0)=y_0$, it leads to
    \begin{equation*}
        y(t)=\frac{p(\bar{x})}{q(\bar{x})}+\left(y_0-\frac{p(\bar{x})}{q(\bar{x})}\right)e^{-q(\bar{x})t}
    \end{equation*}
    and $\lim_{t \to \infty}y(t)=\frac{p(\bar{x})}{q(\bar{x})}$. Now for any system (\ref{dx=f}) with $\lim_{t \to \infty}y^{\mathtt{1}}(t)=\bar{y}^{\mathtt{1}}$, the coupled system
    \begin{equation*}
        \begin{cases}
            \dot{x}^{\mathtt{1}}=f^{\mathtt{1}}(x^{\mathtt{1}},x) \\
            \dot{x}=g^{\mathtt{1}}(x^\mathtt{1},x) \\
            \dot{y}=p(x)-q(x)y
        \end{cases}\begin{array}{l}
        x^{\mathtt{1}}(0)=x^{\mathtt{1}}_0,~x(0)=y^{\mathtt{1}}_0,\\
        y(0)=y_0.
        \end{array}
    \end{equation*}
    supports $\lim_{t \to \infty}x(t)=\bar{x}$, and has the solution
    $$y(t)=e^{-\int_0^tq(s)\mathrm{d}s}\left(y_0+\int_0^{t}p(s)e^{\int_0^sq(u)\mathrm{d}u}\mathrm{d}s\right),$$ which satisfies
    \begin{align*}
        \lim_{t \to \infty}y(t)&=\lim_{t \to \infty}\frac{\int_{0}^tp(x(s))e^{\int_{0}^sq(x(u))\mathrm{d}u}\mathrm{d}s}{e^{\int_{0}^tq(x(s))\mathrm{d}s}} \\
        &=\lim_{t \to \infty}\frac{p(x(t))e^{\int_{0}^tq(x(u))\mathrm{d}u}}{q(x(t))e^{\int_{0}^tq(x(s))\mathrm{d}s}} \\
        &=\lim_{t \to \infty}\frac{p(x(t))}{q(x(t))} \\
        &=\frac{p(\bar{x})}{q(\bar{x})}
    \end{align*}
    with the second equality guaranteed by the L'Hopital's Rule. By definition \ref{Def_dynamic_composable}, the proof is complete.
\end{proof}

\noindent \textbf{Proof of Theorem \ref{Thm_ISS_net}: }

\begin{proof}
    For the coupled system (\ref{coupled_system}), since $\lim_{t \to \infty}y^{\mathtt{1}}(t)=\bar{y}^{\mathtt{1}}$, we just need to prove that $\lim_{t \to \infty}y^{\mathtt{2}}(t)=\bar{y}^{\mathtt{2}}$. Since (\ref{g2_y}) is ISS regarding $(\bar{y}^{\mathtt{1}},\bar{y}^{\mathtt{2}})$, it satisfies
    \begin{equation}\label{ISS_s_to_t}
    \begin{split}
        \left \| y^{\mathtt{2}} \left( t \right) -\bar{y}^{\mathtt{2}} \right \| \le& \beta \left( \left\| y^{\mathtt{2}}(s) - \bar{y}^{\mathtt{2}} \right\| , t-s \right) \\
        +& \gamma ( \sup \limits_{s\le \tau \le t}\left\| x^{\mathtt{2}} \left( \tau \right) -\bar{y}^{\mathtt{1}} \right\| ),
    \end{split}
    \end{equation}
    where $0\le s\le t$, $\beta$ is a class $\mathcal{KL}$ function and $\gamma$ is a class $\mathcal{K}$ function. Substituting $s=\frac{t}{2}$ into (\ref{ISS_s_to_t}) yields
    \begin{equation}\label{ISS_t_2}
    \begin{split}
        \left \| y^{\mathtt{2}} \left( t \right) -\bar{y}^{\mathtt{2}} \right \| \le& \beta \left( \left\| y^{\mathtt{2}}\left(\frac{t}{2}\right) - \bar{y}^{\mathtt{2}} \right\| , \frac{t}{2} \right) \\
        +& \gamma \left( \sup \limits_{\frac{t}{2}\le \tau \le t}\left\| x^{\mathtt{2}} \left( \tau \right) -\bar{y}^{\mathtt{1}} \right\| \right),
    \end{split}
    \end{equation}
    To estimate $\left\| y^{\mathtt{2}}\left(\frac{t}{2}\right) - \bar{y}^{\mathtt{2}} \right\|$, apply (\ref{ISS_s_to_t}) with $s=0$ and $t$ replaced by $\frac{t}{2}$ to obtain
    \begin{equation*}
        \begin{split}
        \left \| y^{\mathtt{2}} \left( \frac{t}{2} \right) -\bar{y}^{\mathtt{2}} \right \| \le& \beta \left( \left\| y^{\mathtt{2}}_0 - \bar{y}^{\mathtt{2}} \right\| , \frac{t}{2} \right) \\
        +& \gamma \left( \sup \limits_{0\le \tau \le \frac{t}{2}}\left\| x^{\mathtt{2}} \left( \tau \right) -\bar{y}^{\mathtt{1}} \right\| \right),
        \end{split}
    \end{equation*}
    Since $\lim_{t\rightarrow \infty} x^{\mathtt{2}}\left( t \right) =\bar{y}^{\mathtt{1}}$, $x^{\mathtt{2}}(t)-\bar{y}^{\mathtt{1}}$ is bounded, then there exists a $M>0$ satisfying $\gamma(\left \| x^{\mathtt{2}}(t)-\bar{y}^{\mathtt{1}} \right \|) \le M$. And for sufficiently large $t$, it holds $\beta \left (\left \| y^{\mathtt{2}}_0-\bar{y}^{\mathtt{2}} \right \|,\frac{t}{2} \right ) \le M$. Substitute all above results into (\ref{ISS_t_2}) to obtain
    \begin{align*}
    \left \| y^{\mathtt{2}} \left( t \right) -\bar{y}^{\mathtt{2}} \right \| \le &\beta \left(  2M , \frac{t}{2} \right) + \gamma \left( \sup_{\frac{t}{2} \le \tau \le t}\left\| x^{\mathtt{2}} \left( \tau \right) -\bar{y}^{\mathtt{1}} \right\| \right) \\
     \rightarrow & ~0 \quad \left( t \rightarrow \infty \right)
    \end{align*}
    It means $y^{\mathtt{2}}(t)\rightarrow \bar{y}^{\mathtt{2}}~ (t \rightarrow \infty)$, the proof is complete.
\end{proof}

\noindent \textbf{Proof of Corollary \ref{coro_thmISS}: }

\begin{proof}
    The proof proceeds by induction on $p$. The result for $p=2$ is described as \textit{Theorem} \ref{Thm_ISS_net} before. Suppose that the result holds for $p=r (r \ge 2)$. Consider $\mathscr{C}^{\mathtt{r}\circ \cdots \circ \mathtt{1}}$ and $\mathscr{C}^{\mathtt{r+1}}$, which satisfy the condition in \textit{Theorem} \ref{Thm_ISS_net}. Then they are dynamically composable, and the proof is complete.
\end{proof}

\noindent \textbf{Proof of Proposition \ref{iss_condition}: }

\begin{proof}
We introduce the following two lemmas for convenience of the subsequent proof: 
\begin{lemma}\label{lem1}(Khalil \cite{Khalil_2002_nonlinear})
    The system (\ref{general_system}) is ISS regarding $(\bar{u},\bar{s})$ if it admits an ISS-Lyapunov function $V: \mathbb{R}^{n-m} \to \mathbb{R}_{\ge 0}$, which is smooth, positive definite, and radially unbounded, and also satisfies that
    \begin{itemize}
        \item [] for bounded input $u$, there exist a positive definite function $W:\mathbb{R}^{n-m} \to \mathbb{R}_{\ge 0}$ and a class $\mathcal{K}$ function $\rho:\mathbb{R}_{\ge 0} \to \mathbb{R}_{\ge 0}$ to support
         \begin{equation*}
            \nabla V^{\top} (s-\bar{s})f(u,s) \le -W(s-\bar{s})
    \end{equation*}
    under the condition of $\left \|s-\bar{s} \right  \| \ge \rho (\left\|u-\bar{u} \right \|)$.
    \end{itemize}
\end{lemma}

\begin{lemma}\label{lem2}(Khalil \cite{Khalil_2002_nonlinear})
    Suppose $f(u,s)$ in (\ref{general_system}) is continuously differentiable and globally Lipschitz in $(u,s)$, and $\bar{s}$ is a globally exponentially stable equilibrium point in (\ref{general_system}) evaluated at $u=\bar{u}$. Then system (\ref{general_system}) is ISS regarding $(\bar{u},\bar{s})$.
\end{lemma}

The proof is obtained by applying \textit{Theorem} \ref{Thm_ISS_net}, \textit{Lemma} \ref{lem1} and \textit{Lemma} \ref{lem2}.
\end{proof}

\noindent \textbf{Proof of Theorem \ref{thm_error_n2}: }

\begin{proof}
    (1) Note that
    \begin{align}\label{esti_0}
        \left| z\left( T \right) -\frac{1}{1+e^{-x_0}} \right|&\le \left| z\left( T \right) -\frac{1}{1+y(\frac{T}{2})} \right| \nonumber\\
        & + \left| \frac{1}{1+y(\frac{T}{2})}- \frac{1}{1+e^{-x_0}} \right|
    \end{align}
    For the first phase of (\ref{sigmoid_2_CRC}), where $0 < t < \frac{T}{2}$, it holds $y(\frac{T}{2})=e^{-x_0\left(1-e^{-\frac{T}{2}}\right)}$ and $z(\frac{T}{2})=z(0)=z_0$. For the second phase of (\ref{sigmoid_2_CRC}), where $\frac{T}{2}< t < T$, it holds $z(T)=\frac{1}{1+y(\frac{T}{2})}+\left( z(\frac{T}{2})-\frac{1}{1+y(\frac{T}{2})} \right)e^{-(1+y(\frac{T}{2}))\frac{T}{2}}$. Then we have
    \begin{flalign}\label{esti_1}
        \left| z\left( T \right) -\frac{1}{1+ y\left( \frac{T}{2} \right)} \right| \nonumber&= \left| z_0-\frac{1}{1+ y\left( \frac{T}{2} \right)} \right|e^{-\left(1+y(\frac{T}{2})\right)\frac{T}{2}}  \nonumber \\
        &\le \left(z_0+1 \right)e^{-\frac{T}{2}}
    \end{flalign}
    and 
    \begin{flalign}
    &\left| \frac{1}{1+y(\frac{T}{2})}- \frac{1}{1+e^{-x_0}} \right| \nonumber \\
    =& \frac{1}{\left (1+y(\frac{T}{2}) \right )(1+e^{-x_0})} \left|y\left(\frac{T}{2}\right)-e^{-x_0} \right| \nonumber \\
    \le& \frac{1}{1+e^{-x_0}} \left|e^{-x_0\left(1-e^{-\frac{T}{2}}\right)}-e^{-x_0} \right| \nonumber \\
    \le& \frac{e^{-x_0}}{1+e^{-x_0}} \left|e^{x_0e^{-\frac{T}{2}}}-1 \right| \nonumber \\
    \le& 2x_0e^{-\frac{T}{2}} \nonumber 
    \end{flalign}
    with the last inequality satisfied for sufficiently large $T\ge T_0>0$. Defining 
    $$C=2x_0+\max_{0\le t\le T_0}\left| \frac{1}{1+y(\frac{t}{2})}- \frac{1}{1+e^{-x_0}} \right|e^{\frac{T_0}{2}},$$
    we have
    \begin{equation}\label{esti_2}
    \left| \frac{1}{1+y(\frac{T}{2})}- \frac{1}{1+e^{-x_0}} \right| \le Ce^{-\frac{T}{2}} 
    \end{equation}
    for all $T>0$. Substitute (\ref{esti_1}) and (\ref{esti_2}) into  (\ref{esti_0}), we can get (\ref{2_step_esti}) with $C_1=z_0+1+C$.

    (2) The composition of msCRC (\ref{exp_CRC}) and (\ref{laycom_sigmoid}) has dynamics of
    \begin{equation*}
        \begin{cases}
        \dot{x}=-x \\
        \dot{y}=-xy \\
        \dot{z}=1-(1+y)z \\
        \end{cases}
    \end{equation*}
    where $y(0)=1$. It is obvious that $y(t)=e^{-x_0 \left(1-e^{-t}\right)}$. Let $u(t)=z(t)-\frac{1}{1+e^{-x_0}}$, and it takes to
    \begin{equation*}
        \dot{u}=-p(t)u+g(t), \quad u(0)=z_0-\frac{1}{1+e^{-x_0}}
    \end{equation*}
    which have solutions
    \begin{align}\label{esti_00}
    u\left( T \right) = & u\left( 0 \right) e^{-\int_0^{T}{p\left( s \right)}\mathrm{d}s} \nonumber\\
    +&e^{-\int_0^{T}{p\left( s \right)}\mathrm{d}s}\int_0^{T}{g\left( s \right) e}^{\int_0^s{p\left( u \right)}\mathrm{d}u}\mathrm{d}s
    \end{align}
    where
    \begin{align*}
    p(t) &=1+y(t)=1+e^{-x_0 \left(1-e^{-t}\right)}, \\
    g(t) &=1-\frac{1}{1+e^{-x_0}}(1+y(t)) \\
    &=\frac{1}{e^{x_0}+1}\left(1-e^{x_0e^{-t}} \right).
    \end{align*}
    Then it holds that
    \begin{align}\label{esti_4}
        \left| u(0)e^{-\int_0^{T}{p(s)\mathrm{d}s}} \right|&=\left| u(0)e^{-\int_0^{T}{\left( 1+e^{-x_0 \left(1-e^{-s}\right)}\right )\mathrm{d}s}} \right| \nonumber \\
    &\le \left| u\left( 0 \right) \right|e^{-T}=C_3e^{-T}
    \end{align}
    Besides, defining $F(t)=\left| e^{-\int_0^{t}{p\left( s \right) \mathrm{d}s}}\int_0^{t}{g\left( s \right) e^{\int_0^s{p\left( u \right) \mathrm{d}u}}\mathrm{d}s} \right|$ and using L'Hopital's Rule, we have
    \begin{align*}
    \lambda&=-\lim_{t \to \infty} \frac{\ln F(t)}{t}   \\ 
    &=-\lim_{t\rightarrow \infty}\frac{\ln \left( e^{-\int_0^{t}{p\left( s \right) \mathrm{d}s}}\int_0^{t}{(-g\left( s \right) )e^{\int_0^s{p\left( u \right) \mathrm{d}u}}\mathrm{d}s} \right)}{t}\\
    &=-\lim_{t\rightarrow \infty}\left( \frac{-\int_0^{t}{p\left( s \right) \mathrm{d}s}}{t}+\frac{\ln \left( -\int_0^{t}{g\left( s \right) e^{\int_0^s{p\left( u \right) \mathrm{d}u}}\mathrm{d}s} \right)}{t} \right) \\
    &=-\lim_{t\rightarrow \infty}\left( -p\left( t \right) +\frac{g\left( t \right) e^{\int_0^{t}{p\left( u \right) \mathrm{d}u}}}{\int_0^{t}{g\left( s \right) e^{\int_0^s{p\left( u \right) \mathrm{d}u}}\mathrm{d}s}} \right) \\
    &=1+e^{-x_0}-\lim_{t\rightarrow \infty}\left( \frac{g^{\prime}(t)}{g( t )}+p(t) \right) \\
    &=-\lim_{t\rightarrow \infty}\frac{x_0e^{x_0e^{-t}-t}}{1-e^{x_0e^{-t}}}\\
    &=-\lim_{t\rightarrow \infty}\frac{-x_0e^{-t}}{x_0e^{x_0e^{-t}}e^{-t}}\\
    &=1.\\
\end{align*}
Then for any $\varepsilon >0$, there is a $T_0>0$ such that for any $T\ge T_0$, it holds 
\begin{equation*}
    -\frac{\ln F(T)}{T}\ge  \lambda-\varepsilon.
\end{equation*}
Hence we have $F(T) \le e^{-(1-\varepsilon)T},~ T\ge T_0$. Defining 
$$C_2=1+\max_{0\le t\le T_0}F(t)e^{(1-\varepsilon)T_0},$$
we have
\begin{equation}\label{esti_6}
    \left |e^{-\int_0^{T}{p\left( s \right)}\mathrm{d}s}\int_0^{T}{g\left( s \right) e}^{\int_0^s{p\left( u \right)}\mathrm{d}u}\mathrm{d}s \right |=F(T) \le C_2 e^{-(1-\varepsilon)T}
\end{equation}
for all $T\ge 0$. Substitute (\ref{esti_4}) and (\ref{esti_6}) into (\ref{esti_00}), we can get (\ref{1_step_esti}).
\end{proof}

\subsection{Another example of composite function computed}\label{appendixB}
\setcounter{equation}{0}
\renewcommand\theequation{B.\arabic{equation}}

Consider the frequently-used quadratic formula (positive root) 
\begin{equation}\label{root}
    z=\frac{1}{2}\left (b+\sqrt{b^2+4c}\right )
\end{equation}
for the quadratic equation $z^2-bz-c=0$ with $b,c>0$. This composition function may be viewed as a composition of three functions $x=b^2+4c$, $y=\sqrt{x}$ and $z=\frac{1}{2}(b+y)$. We thus introduce the following three msCRCs
\begin{equation*}
    \begin{split}
        \mathscr{C}^{\mathtt{1}}:&~C \overset{4}{\longrightarrow}C+X, \quad X \overset{1}{\longrightarrow} \varnothing, \quad 2B \overset{1}{\longrightarrow}2B+X;    \\
        \mathscr{C}^{\mathtt{2}}:&~X \overset{1}{\longrightarrow} X+Y,\quad  2Y \overset{1}{\longrightarrow} \varnothing; \\
        \mathscr{C}^{\mathtt{3}}:&~ B \overset{1}{\longrightarrow}B+Z,\quad Y \overset{1}{\longrightarrow} Y +Z,\quad  Z \overset{2}{\longrightarrow} \varnothing.
    \end{split}
\end{equation*}
Then the dynamics of the composition $\mathscr{C}^{\mathtt{3}\circ \mathtt{2}\circ \mathtt{1}}$ is
\begin{equation*}
    \begin{cases}
        \dot{b}=\dot{c}=0, \\
        \dot{x}=b^2+4c-x, \\
        \dot{y}=x-y^2, \\
        \dot{z}=b+y-2z.
    \end{cases}\begin{array}{ll}
         b(0)=b_0,&c(0)=c_0, \\
         x(0)=x_0,&y(0)=y_{0}, \\
         z(0)=z_0.
    \end{array}
\end{equation*}
It is easy to observe that $\bar{x}=b_0^2+4c_0,~\bar{y}=\sqrt{b_0^2+4c_0}, ~\bar{z}=\frac{1}{2}\left(b_0+\sqrt{b_0^2+4c_0}\right)$. Next, construct $V(y-\bar{y})=\frac{1}{2}(y-\bar{y})^2$, which suggests
\begin{equation*}
    \begin{split}
        &\frac{\mathrm{d}V}{\mathrm{d}t} =(y-\bar{y})\left(x-y^2 \right) \\
        =&-(y+\bar{y})(y-\bar{y})^2+(x-\bar{x})(y-\bar{y}) \\
        \le &-2\bar{y}V+|y-\bar{y}||x-\bar{x}|. \\
    \end{split}
\end{equation*}
Since $y(t)$ is uniformly bounded in time when $x(t)$ are bounded (suggested by $\mathscr{C}^{\mathtt{1}}$), there is a $C>0$ such that $|y-\bar{y}|<C$. We can thus get $\frac{\mathrm{d}V}{\mathrm{d}t}\le -2\bar{y}V+C|x-\bar{x}|$. Similar to the proof in \textit{Example} \ref{ex_sqrt}, we can conclude that $\dot{y}=x-y^2$ is ISS regarding $\bar{x}$. Further, $\dot{z}=b+y-z$ is ISS regarding $(\bar{b},\bar{y})$ by \textit{Proposition} \ref{prop_simple_structure}. Then by \textit{Corollary} \ref{coro_thm1} and \textit{Corollary} \ref{coro_thmISS}, $\mathscr{C}^{\mathtt{1}}$ and $\mathscr{C}^{\mathtt{2}}$, $\mathscr{C}^{\mathtt{2}\circ \mathtt{1}}$ and $\mathscr{C}^{\mathtt{3}}$ are both dynamically composable, and $\mathscr{C}^{\mathtt{3} \circ \mathtt{2}\circ \mathtt{1}}$ serves to dynamically compute (\ref{root}).

\section*{References}

\bibliographystyle{ieeetr}
\bibliography{main}

\begin{thebibliography}{10}

\bibitem{Chen_2024_synthetic}
Z.~Chen, J.~M. Linton, S.~Xia, X.~Fan, D.~Yu, J.~Wang, R.~Zhu, and M.~B. Elowitz, ``A synthetic protein-level neural network in mammalian cells,'' {\em Science}, vol.~386, no.~6727, pp.~1243--1250, 2024.

\bibitem{Danino_2015_programmable}
T.~Danino, A.~Prindle, G.~A. Kwong, M.~Skalak, H.~Li, K.~Allen, J.~Hasty, and S.~N. Bhatia, ``Programmable probiotics for detection of cancer in urine,'' {\em Science Translational Medicine}, vol.~7, no.~289, pp.~289ra84--289ra84, 2015.

\bibitem{Peralta_2012_microbial}
P.~P. Peralta-Yahya, F.~Zhang, S.~B. Del~Cardayre, and J.~D. Keasling, ``Microbial engineering for the production of advanced biofuels,'' {\em Nature}, vol.~488, no.~7411, pp.~320--328, 2012.

\bibitem{Feinberg_1987_chemical}
M.~Feinberg, ``Chemical reaction network structure and the stability of complex isothermal reactors—i. the deficiency zero and deficiency one theorems,'' {\em Chemical Engineering Science}, vol.~42, no.~10, pp.~2229--2268, 1987.

\bibitem{Craciun_2013_persistence}
G.~Craciun, F.~Nazarov, and C.~Pantea, ``Persistence and permanence of mass-action and power-law dynamical systems,'' {\em SIAM Journal on Applied Mathematics}, vol.~73, no.~1, pp.~305--329, 2013.

\bibitem{Domijan_2009_bistability}
M.~Domijan and M.~Kirkilionis, ``Bistability and oscillations in chemical reaction networks,'' {\em Journal of Mathematical Biology}, vol.~59, no.~4, pp.~467--501, 2009.

\bibitem{Conradi_2007_saddle}
C.~Conradi, D.~Flockerzi, and J.~Raisch, ``Saddle-node bifurcations in biochemical reaction networks with mass action kinetics and application to a double-phosphorylation mechanism,'' in {\em 2007 American Control Conference}, pp.~6103--6109, IEEE, 2007.

\bibitem{Adleman_1994_molecular}
L.~M. Adleman, ``Molecular computation of solutions to combinatorial problems,'' {\em Science}, vol.~266, no.~5187, pp.~1021--1024, 1994.

\bibitem{Arkin_1994_computational}
A.~Arkin and J.~Ross, ``Computational functions in biochemical reaction networks,'' {\em Biophysical Journal}, vol.~67, no.~2, pp.~560--578, 1994.

\bibitem{Buisman_2009_computing}
H.~Buisman, H.~M. ten Eikelder, P.~A. Hilbers, and A.~M. Liekens, ``Computing algebraic functions with biochemical reaction networks,'' {\em Artificial Life}, vol.~15, no.~1, pp.~5--19, 2009.

\bibitem{Fages_2017_strong}
F.~Fages, G.~Le~Guludec, O.~Bournez, and A.~Pouly, ``Strong turing completeness of continuous chemical reaction networks and compilation of mixed analog-digital programs,'' in {\em Computational Methods in Systems Biology: 15th International Conference, CMSB 2017, Darmstadt, Germany, September 27--29, 2017, Proceedings 15}, pp.~108--127, Springer, 2017.

\bibitem{Soloveichik_2010_dna}
D.~Soloveichik, G.~Seelig, and E.~Winfree, ``Dna as a universal substrate for chemical kinetics,'' {\em Proceedings of the National Academy of Sciences}, vol.~107, no.~12, pp.~5393--5398, 2010.

\bibitem{Cherry_2018_scaling}
K.~M. Cherry and L.~Qian, ``Scaling up molecular pattern recognition with dna-based winner-take-all neural networks,'' {\em Nature}, vol.~559, no.~7714, pp.~370--376, 2018.

\bibitem{Okumura_2022_nonlinear}
S.~Okumura, G.~Gines, N.~Lobato-Dauzier, A.~Baccouche, R.~Deteix, T.~Fujii, Y.~Rondelez, and A.~J. Genot, ``Nonlinear decision-making with enzymatic neural networks,'' {\em Nature}, vol.~610, no.~7932, pp.~496--501, 2022.

\bibitem{Vasic_2020_crn++}
M.~Vasi{\'c}, D.~Soloveichik, and S.~Khurshid, ``Crn++: Molecular programming language,'' {\em Natural Computing}, vol.~19, no.~2, pp.~391--407, 2020.

\bibitem{Chou_2017_chemical}
C.~T. Chou, ``Chemical reaction networks for computing logarithm,'' {\em Synthetic Biology}, vol.~2, no.~1, p.~ysx002, 2017.

\bibitem{Chen_2014_rate}
H.-L. Chen, D.~Doty, and D.~Soloveichik, ``Rate-independent computation in continuous chemical reaction networks,'' in {\em Proceedings of the 5th Conference on Innovations in Theoretical Computer Science}, pp.~313--326, 2014.

\bibitem{Chalk_2019_composable}
C.~Chalk, N.~Kornerup, W.~Reeves, and D.~Soloveichik, ``Composable rate-independent computation in continuous chemical reaction networks,'' {\em IEEE/ACM Transactions on Computational Biology and Bioinformatics}, vol.~18, no.~1, pp.~250--260, 2019.

\bibitem{Samaniego_2020_sequestration}
C.~C. Samaniego, J.~Kim, and E.~Franco, ``Sequestration and delays enable the synthesis of a molecular derivative operator,'' in {\em 2020 59th IEEE Conference on Decision and Control (CDC)}, pp.~5106--5112, IEEE, 2020.

\bibitem{Anderson_2025_chemical}
D.~F. Anderson and B.~Joshi, ``Chemical mass-action systems as analog computers: Implementing arithmetic computations at specified speed,'' {\em Theoretical Computer Science}, vol.~1025, p.~114983, 2025.

\bibitem{Vasic_2022_programming}
M.~Vasi{\'c}, C.~Chalk, A.~Luchsinger, S.~Khurshid, and D.~Soloveichik, ``Programming and training rate-independent chemical reaction networks,'' {\em Proceedings of the National Academy of Sciences}, vol.~119, no.~24, p.~e2111552119, 2022.

\bibitem{Anderson_2021_reaction}
D.~F. Anderson, B.~Joshi, and A.~Deshpande, ``On reaction network implementations of neural networks,'' {\em Journal of the Royal Society Interface}, vol.~18, no.~177, p.~20210031, 2021.

\bibitem{Fan_2025_automatic}
Y.~Fan, X.~Zhang, C.~Gao, and D.~Dochain, ``Automatic implementation of neural networks through reaction networks—part i: Circuit design and convergence analysis,'' {\em IEEE Transactions on Automatic Control}, 2025.

\bibitem{Lachmann_1995_computationally}
M.~Lachmann and G.~Sella, ``The computationally complete ant colony: Global coordination in a system with no hierarchy,'' in {\em European Conference on Artificial Life}, pp.~784--800, Springer, 1995.

\bibitem{Shi_2025_controlling}
X.~Shi, C.~Gao, and D.~Dochain, ``Controlling the occurrence sequence of reaction modules through biochemical relaxation oscillators,'' {\em Automatica}, vol.~176, p.~112261, 2025.

\bibitem{Fan_2024_automatic}
Y.~Fan, X.~Zhang, C.~Gao, and D.~Dochain, ``Automatic implementation of neural networks through reaction networks--part ii: Error analysis,'' {\em arXiv preprint arXiv:2401.07077}, 2024.

\bibitem{Sontag_1989_smooth}
E.~D. Sontag, ``Smooth stabilization implies coprime factorization,'' {\em IEEE Transactions on Automatic Control}, vol.~34, no.~4, pp.~435--443, 1989.

\bibitem{Rouche_1977_stability}
N.~Rouche, P.~Habets, and M.~Laloy, {\em Stability theory by Liapunov's direct method}, vol.~4.
\newblock Springer, 1977.

\bibitem{Khalil_2002_nonlinear}
H.~K. Khalil and J.~W. Grizzle, {\em Nonlinear Systems}, vol.~3.
\newblock Prentice hall Upper Saddle River, NJ, 2002.

\end{thebibliography}

\end{document}